\documentclass{elsarticle}

\usepackage{graphicx}
\usepackage{amsmath}
\usepackage{amssymb}
\usepackage{multirow}
\usepackage{array}

\newcommand{\tensor}[1]{\vec{\vec{#1}}}
\newcommand{\sub}[1]{_{\textrm{#1}}} 
\newcommand{\super}[1]{^{\textrm{#1}}} 
\newcommand{\sci}[2]{\ensuremath{#1 \times 10^{#2}}}

\title{Efficient classical density-functional theories of rigid-molecular fluids
	and a simplified free energy functional for liquid water}
\author{Ravishankar Sundararaman}
\author{T. A. Arias}
\address{Department of Physics, Cornell University, Ithaca, NY 14853, USA}

\begin{document}
\begin{abstract}

Classical density-functional theory provides an efficient alternative to molecular dynamics
simulations for understanding the equilibrium properties of inhomogeneous fluids.
However, application of density-functional theory to multi-site molecular fluids
has so far been limited by complications due to the implicit molecular geometry constraints
on the site densities, whose resolution typically requires expensive Monte Carlo methods.
Here, we present a general scheme of circumventing this so-called inversion problem:
compressed representations of the orientation density.
This approach allows us to combine the superior iterative convergence properties of multipole
representations of the fluid configuration with the improved accuracy of site-density functionals.
Next, from a computational perspective, we show how to extend the DFT++ algebraic formulation of
electronic density-functional theory to the classical fluid case and present a basis-independent
discretization of our formulation for molecular classical density-functional theory.
Finally, armed with the above general framework, we construct a simplified free-energy functional
for water which captures the radial distributions, cavitation energies, and the linear and non-linear
dielectric response of liquid water.  The resulting approach will enable efficient and reliable
first-principles studies of atomic-scale processes in contact with solution or other liquid environments.
\end{abstract}

\maketitle


\section{Introduction} \label{sec:Intro}

The microscopic structure of liquids plays an important role in several
biological processes and chemical systems of technological importance,
and is the subject of continued scientific interest. 
Several computational techniques have been developed to study 
the bulk and inhomogeneous properties of liquids.
(See \cite{WhatIsLiquid} for a comprehensive review.)

Monte Carlo calculations and molecular dynamics simulations with a 
simplified Hamiltonian, often composed of additive pair-potentials,
are the most popular techniques used to compute properties of 
inhomogeneous liquids. However these can be quite expensive due to
the long equilibration times and extensive phase-space sampling
necessary to compute thermodynamic averages with sufficiently low
statistical noise.

Theories in terms of the equilibrium densities rather than individual
configurations of molecules avoid this phase space sampling and hence are
much more efficient for the computation of equilibrium properties. 
Integral equation theories, based on approximations to the diagrammatic series
of interactions, can be reasonably accurate but still prove relatively expensive
and have only recently been applied to inhomogeneous systems
in three dimensions \cite{IntEqnInhomogeneous}.

All of the above methods require the construction of a simplified Hamiltonian,
usually restricted to pair potentials. Many applications, such as the determination
of chemical reaction pathways, also require estimation of free energies,
which involves a coupling constant integration and hence incurs additional costs.
Classical density-functional theories based on an exact variational theorem for the free energy
of a liquid \cite{ThermalDFT-Mermin} avoid these restrictions, at least in principle.
In practice, they involve directly approximating the free energy as a functional
of the liquid density. They have the further advantage of being readily coupled
to a quantum mechanical calculation of an electronic system within the framework
of joint density-functional theory \cite{JDFT}, which makes quantum treatment
practical for much larger systems than possible with
\emph{ab initio} molecular dynamics \cite{CPMD}.

Free-energy functional approximations for fluids of spherical particles often
employ a thermodynamic perturbation about the hard sphere fluid described
accurately by fundamental measure theory \cite{RosenfeldFMT,FMTreview}.
These may be extended to model polar fluids such as the Stockmeyer fluid \cite{DipoleFluid},
but the accuracy of such theories for real molecular fluids is not satisfactory.

Molecular fluids are best described within the reduced interaction-site models (RISM)
\cite{RISM1}, which express the interactions in terms of a few sites on each molecule,
usually on atomic centers constrained by a rigid model molecular geometry.
The free energy functional descriptions in terms of these site densities, however, is
complicated by the molecular geometry constraints; even the ideal-gas free-energy is no longer
expressible as an analytical closed-form functional of the site-densities alone.
An explicit functional can be written by
introducing effective ideal-gas site potentials as auxiliary variables \cite{RISM2},
but this still requires inversion of an integral equation to obtain these potentials from the
site densities, a problem which can be solved explicitly only in some limits such as reducing the
molecule to a point \cite{WaterFreezingDFT},
and requires an expensive Monte Carlo integration for the general case.

The above inversion problem can effectively be avoided \cite{LischnerHCl} by switching 
to the site potentials as the independent variables \emph{instead} of the site densities.
This method was applied successfully to fluids of hydrogen chloride
\cite{LischnerHCl} and water \cite{LischnerH2O} in one dimension. The convergence of
free energy minimization with respect to these independent variables turns out to be quite slow,
however, particularly in the presence of strong electric fields. 

This work presents a simple general scheme of choosing independent variables that
can generate the site densities for the free-energy functional treatment of
molecular fluids. In Section~\ref{sec:IdealGas}, we demonstrate the site-potential
solution as a special case of this general scheme and present other representations
with better iterative convergence during free energy minimization.
In section~\ref{sec:ExcessFunctionals}, we construct a simplified semi-empirical
excess functional for liquid water which adequately captures the properties
most critical to successful \emph{ab initio} treatment of solvation within 
the framework of joint density-functional theory.
Finally in section~\ref{sec:Results}, we detail the computational implementation
of the above theories in the open-source plane-wave density-functional theory software JDFTx
\cite{JDFTx}, using the basis-independent DFT++ algebraic formulation \cite{AlgebraicDFT},
and present numerical studies of the molecular classical density-functional framework
and the free energy functional for liquid water.

\section{Free energy of an ideal gas of rigid molecules} \label{sec:IdealGas}

The site-density-functional theory of molecular fluids is based
on functional approximations to the in-principle exact free-energy
functional
\begin{equation}
\Phi[\{N_\alpha(\vec{r})\}] = \Phi\sub{id}[\{N_\alpha\}] + F\sub{ex}[\{N_\alpha\}],
\label{eqn:ExactFunctional}
\end{equation}
where $\Phi$ is the grand free energy of the interacting fluid, $\Phi\sub{id}$
is the exact grand free energy for the molecular ideal gas, $N_\alpha(\vec{r})$ are the 
densities of distinct sites (indexed by $\alpha$) in the molecule, and $F\sub{ex}$
captures the effect of all the interactions and correlations. The equilibrium densities
and free energy are obtained by minimizing the free energy over all allowed densities.

The heart of the inversion problem lies in the fact that the site densities $N_\alpha(\vec{r})$
are not independent variables, but are constrained by the assumption of a rigid-molecular geometry.
For definiteness, let the molecule geometry be specified by $\vec{R}_{\alpha k}$, the positions of the
sites for a molecule centered at the origin in some reference orientation. Here, $\alpha$ indexes
distinct sites while $k$ indexes multiple sites of the same type equivalent under the symmetry
of the molecule (e.g. for a 3-site water model, $\alpha\in\{O,H\}$, $k=1$ for $\alpha=O$ and 
$k\in\{1,2\}$ for $\alpha=H$.)

\subsection{Treatment of site-density constraints}

The inversion problem in the original approach of \cite{RISM2,WaterFreezingDFT},
which includes ideal-gas effective potentials $\psi_\alpha(\vec{r})$
as auxiliary independent variables in addition to the site densities,
is avoided in \cite{LischnerHCl} by switching to $\psi_\alpha(\vec{r})$
as the sole independent variables. The site densities and ideal-gas free energy
in the presence of external site potentials $V_\alpha$ and chemical potentials $\mu_\alpha$
are then expressed in terms of the $\psi_\alpha(\vec{r})$ using
\begin{flalign}
\Phi\sub{id}[\{\psi_\alpha(\vec{r})\}] &= \Omega^{(ni)}[\{\psi_\alpha\}]
	+ \sum_\alpha \int \textrm{d}\vec{r} N_\alpha(\vec{r}) (V_\alpha(\vec{r}) - \mu_\alpha - \psi_\alpha(\vec{r})) \\
\Omega^{(ni)} &\equiv -N\sub{ref}T \int\prod_{\alpha,k} e^{-\psi_\alpha(\vec{r}_{\alpha k})/T}
	\textrm{d}\vec{r}_{\alpha k} s(\{\vec{r}_{\alpha k}\}) \label{eqn:OmegaNI}\\
N_\alpha(\vec{r}) &\equiv \frac{\delta\Omega^{(ni)}}{\delta\psi_\alpha(\vec{r})}, \label{eqn:NalphaFromOmegaNI}
\end{flalign}
Here, the reference density $N\sub{ref}$ sets the zero of chemical potential
and the constraint function $s(\{\vec{r}_{\alpha k}\})$ picks out configurations
$\{\vec{r}_{\alpha k}\}$ which satisfy the rigid molecule geometry
(i.e. equivalent to $\{\vec{R}_{\alpha k}\}$ under rotations and translations).

Employing a spherical harmonic expansion of the constraint function,
\cite{LischnerHCl} and \cite{LischnerH2O} specialize (\ref{eqn:OmegaNI})
for diatomic and triatomic molecules respectively.
However, that expansion also becomes computationally challenging as one
moves to calculations without planar symmetry.
Instead, we transform (\ref{eqn:OmegaNI}) to
\begin{equation}
\Omega^{(ni)} = -N\sub{ref}T \int \frac{\textrm{d}\vec{r} \textrm{d}\omega}{8\pi^2}
	\prod_{\alpha,k} e^{-\psi_\alpha(\vec{r}+\omega\circ\vec{R}_{\alpha k})/T}
\end{equation}
where $\omega\in SO(3)$ is a rotation and $\omega\circ\vec{R}$
is the result of rotating vector $\vec{R}$ by $\omega$,
and we directly discretize the integral over orientations as described in
\ref{sec:SO3quad} for practical calculations in three dimensions.

It is instructive to further transform the above equations to
\begin{flalign}
\Phi\sub{id} &=
	T \int \frac{\textrm{d}\vec{r} \textrm{d}\omega}{8\pi^2} p_\omega(\vec{r})
		\left(\log\frac{p_\omega(\vec{r})}{N_\textrm{ref}} - 1\right)
	+ \sum_\alpha \int \textrm{d}\vec{r} N_\alpha(\vec{r}) (V_\alpha(\vec{r}) - \mu_\alpha)
	\label{eqn:PhiIDfromPomega} \\
N_\alpha(\vec{r}) &= \sum_{k} \int \frac{\textrm{d}\omega}{8\pi^2} p_\omega\left(\vec{r}
	- \omega\circ\vec{R}_{\alpha k}\right)
	\label{eqn:NalphaFromPomega}
\end{flalign}
with
\begin{equation}
p_\omega(\vec{r}) = N\sub{ref}\prod_{\alpha,k}
	e^{-\psi_\alpha(\vec{r}+\omega\circ\vec{R}_{\alpha k})/T}.
	\label{eqn:PomegaFromPsiAlpha}
\end{equation}
Here, $p_\omega(\vec{r})$ represents the probability of finding a molecule
centered at location $\vec{r}$ with orientation $\omega$. For an ideal molecular gas,
$p_\omega(\vec{r})$ is simply a product of Boltzmann factors for each site 
given that $\psi_\alpha(\vec{r})$ are ideal-gas effective potentials since they
equal $V_\alpha(\vec{r})-\mu_\alpha$ when $\Phi\sub{id}$ is minimized.

Note that, given the explicit
expressions for the ideal-gas free energy (\ref{eqn:PhiIDfromPomega})
and site densities (\ref{eqn:NalphaFromPomega}),  $p_\omega(\vec{r})$
is a natural choice for the independent variables
for \emph{unconstrained} free-energy minimization.
Section \ref{sec:Convergence} demonstrates that conjugate gradients
minimization over $p_\omega(\vec{r})$ as the independent variables converges
much faster than minimization over the $\{\psi_\alpha(\vec{r})\}$.
A potential disadvantage of using $p_\omega(\vec{r})$ is the increased
memory requirement, but practical calculations of reasonable size
are possible using the efficient orientation quadratures of \ref{sec:SO3quad}.
Moreover the superior convergence properties can be retained, 
while mitigating the memory requirements, by switching to \emph{compressed}
multipole representations of $p_\omega(\vec{r})$, as we now discuss.

\subsection{Representations of the Orientation Density} \label{sec:Representations}

Our first task in this development is to demonstrate that
minimizing free energy functionals over $p_\omega(\vec{r})$
yields the same results as minimizing over $\{\psi_\alpha(\vec{r})\}$.  To demonstrate this
we employ a constrained search procedure, to find
\begin{flalign*}
\Phi &= \min_{p_\omega(\vec{r})} \left( \Phi\sub{id}[p_\omega(\vec{r})] + F\sub{ex}[\{N_\alpha\}] \right) \\
	&= \min_{p_\omega(\vec{r})} \bigg( 
		T \underbrace{ \int \frac{\textrm{d}\vec{r} \textrm{d}\omega}{8\pi^2} 
				p_\omega(\vec{r}) \log\frac{p_\omega(\vec{r})}{N_\textrm{ref}} }
			_{-S\sub{id}[p_\omega(\vec{r})]}
		+ F\sub{id-ex}[\{N_\alpha\}] \bigg),
\end{flalign*}
which follows because all the terms in $\Phi\sub{id}$ are explicit site-density functionals
except for the molecular ideal gas entropy ($S\sub{id}$) contribution separated out above. Next, the minimization
over all $p_\omega(\vec{r})$ can be performed by minimizing over those 
$p_\omega(\vec{r})$ that yield a specific set of site densities $\{N_\alpha(\vec{r})\}$,
and then minimizing over all $\{N_\alpha(\vec{r})\}$,
\begin{flalign*}
\Phi &= \min_{\{N_\alpha(\vec{r})\}}\left( \min_{p_\omega(\vec{r})\mapsto\{N_\alpha(\vec{r})\}} 
		T \int \frac{\textrm{d}\vec{r} \textrm{d}\omega}{8\pi^2} p_\omega(\vec{r}) \log\frac{p_\omega(\vec{r})}{N_\textrm{ref}}
		+ F\sub{id-ex}[\{N_\alpha\}]\right).
\end{flalign*}
Finally, the inner, constrained minimization over $p_\omega(\vec{r})$ that lead to given
site densities can be performed explicitly by introducing
Lagrange multipliers $\psi_\alpha(\vec{r})$ for each $N_\alpha(\vec{r})$ constraint.
It is straightforward to verify that the Euler-Lagrange equation for that
extremization is precisely (\ref{eqn:PomegaFromPsiAlpha}), so that the
result of free energy minimization over $p_\omega(\vec{r})$ is
exactly the same as the ideal-gas effective potential methods
of \cite{RISM2,LischnerHCl}.

To generalize this approach, we note that the exact equivalence
between minimization over $p_\omega(\vec{r})$ and minimization
over $\{\psi_\alpha(\vec{r})\}$ holds only when the
external potential takes the form of external site potentials
$V_\alpha(\vec{r})$. In principle, we could go beyond the
reduced-interaction site model and consider arbitrary orientation
dependent external potentials $V_\omega(\vec{r})$
(of which site potentials $V_\alpha(\vec{r})$ are a special case).  
From this perspective, the minimization over $\{\psi_\alpha(\vec{r})\}$
can be reinterpreted as a minimization over only those $p_\omega(\vec{r})$ that
maximize the molecular ideal gas entropy $S\sub{id}[p_\omega(\vec{r})]$
subject to site-density constraints $\{N_\alpha(\vec{r})\}$
(whose Lagrange-multiplier constraints become the site potentials.)
The variational principle implies that this procedure will
always result in a free-energy greater than or equal to direct, unconstrained
minimization over $p_\omega(\vec{r})$, with equality guaranteed only
when the external orientation potential $V_\omega(\vec{r})$ can be
reduced to site potentials $V_\alpha(\vec{r})$.

These considerations lead to the perspective of the $\{\psi_\alpha(\vec{r})\}$
as a \emph{compressed representation} of $p_\omega(\vec{r})$, with
decompression carried out by maximizing the entropy subject to
constraints for which the $\{\psi_\alpha(\vec{r})\}$ are Lagrange multipliers.
From the most general perspective, then, any set of functional constraints
$\{X_i=\hat{X}_i[p_\omega(\vec{r})]\}$ corresponds to a maximum-entropy
compressed representation of $p_\omega(\vec{r})$, where the independent variables $\chi_i$
for the free-energy functional minimization are the Lagrange multipliers for the
corresponding $X_i$ constraint in the maximization of $S\sub{id}[p_\omega(\vec{r})]$.
Specifically,
\begin{equation}
\Phi = \min_{\{\chi_i\}} \left(
	\Phi\sub{id}\left[ p_\omega(\vec{r})[\chi_i] \right]
	+ F\sub{ex}\left[ \{N_\alpha[p_\omega(\vec{r})[\chi_i]]\} \right],
	\right)
\end{equation}
where $p_\omega(\vec{r})[\chi_i]$ is the solution of
\begin{equation}
\frac{\delta}{\delta p_\omega(\vec{r})} \left(
	T S\sub{id}[p_\omega(\vec{r})] + \sum_i (\hat{X}_i[p_\omega(\vec{r})] - X_i) \chi_i
	\right) = 0.
\end{equation}
Here, $\Phi\sub{id}[p_\omega(\vec{r})]$ and $N_\alpha[p_\omega(\vec{r})]$
are given by (\ref{eqn:PhiIDfromPomega}) and (\ref{eqn:NalphaFromPomega}) respectively.
Note that $i$ typically includes a continuous index such as $\vec{r}$,
and $\sum_i$ then denotes the corresponding integrals.

From this new perspective, picking $\hat{X}_i[p_\omega(\vec{r})]=N_\alpha[p_\omega(\vec{r})]$
yields the ideal-gas site-potential representation with $\chi_i = \psi_\alpha(\vec{r})$
as the independent variables and $p_\omega(\vec{r})$ given by (\ref{eqn:PomegaFromPsiAlpha}).
Similarly, picking $\hat{X}_i[p_\omega(\vec{r})] = p_\omega(\vec{r})$ yields the
trivial self-representation, with $p_\omega(\vec{r})$ as the independent variables.
As shown earlier, both these representations are exact when the external potentials
are site potentials, while the former is a variational approximation to the latter
in the most general case of orientation potentials.

The advantage of this general framework is that we can develop new, physically motivated
representations which then are guaranteed to be variational approximations.
Of particular interest are representations based on multipole probability densities
\begin{equation}
\hat{M}^j_{m_1m_2}(\vec{r})[p_\omega(\vec{r})] = \frac{2j+1}{8\pi^2}
	\int \textrm{d}\omega p_\omega(\vec{r}) D^j_{m_1 m_2}(\omega),
\end{equation}
where $D^j_{m_1 m_2}(\omega)$ are the Wigner $D$-matrices
\cite{GroupTheory-Wigner} (irreducible matrix representations of $SO(3)$).
The Lagrange-multiplier independent variables $\mu^j_{m_1m_2}(\vec{r})$
resulting from this choice then generate the orientation probability
\begin{equation}
p_\omega(\vec{r})[\mu^j_{m_1m_2}(\vec{r})] = N\sub{ref} \prod_j \prod^{+j}_{m_1,m_2=-j}
	\exp\left(-\frac{\mu^j_{m_1 m_2}(\vec{r}) D^j_{m_1 m_2}(\omega)}{T} \right). \label{eqn:PomegaFromMujmm}
\end{equation}
By the completeness of the $D^j_{m_1 m_2}$ on $SO(3)$,
this representation is exact if all components $j\to\infty$ are included.
In practice, we truncate the expansion at finite $j$.\footnote{
This expansion in $j$ is different from the spherical harmonic
expansion of $\Omega^{(ni)}$ for triatomic molecules introduced in \cite{LischnerH2O}.
In particular, truncating expansion (\ref{eqn:PomegaFromMujmm}) at $j=1$ retains the exact nonlinear
dielectric response for axisymmetric molecules, whereas the corresponding truncation in
\cite{LischnerH2O} would incur a $20\%$ error in the $O(E^2)$ term of $\epsilon(E)$ at ambient conditions.}

We find below that including terms up to $j=1$ is sufficient
for many practical problems, particularly when the
external potential is dominated by strong electric fields.
We choose to label the corresponding independent variables for this truncation
as $\mu(\vec{r})$ for $j=0$ and $\vec{\epsilon}(\vec{r})$ for $j=1$,
because they correspond to the ideal-gas effective
local chemical potential and local electric field
(up to factors of $T$ and the molecule's dipole moment).
Section~\ref{sec:Convergence} below compares the accuracy and
convergence properties of this $\{\mu,\vec{\epsilon}\}$ representation
to those of the site-potential ($\{\psi_\alpha\}$) representation
and the self-representation ($p_\omega$).

Finally, we would like to point out that this general perspective
opens up a promising avenue for excess functional development.
Our framework enables the computation of site densities
\emph{and} multipole densities irrespective of the independent
variables used for minimization, which facilitates the generalization
of site-density excess functionals $F\sub{ex}[\{N_\alpha\}]$ to combined
site-multipole functionals $F\sub{ex}[\{N_\alpha\}, \{M^j_{m_1m_2}\}]$
or even to full orientation density functionals $F\sub{ex}[p_\omega]$.
In particular, it should now be possible to combine the best features of
site-density functionals, which better capture short-ranged correlations, with those of 
multipole functionals, which allow for analytically derivable long-range correlations.

\section{Excess functionals} \label{sec:ExcessFunctionals}

So far we have focused on accurate and efficient representations
of the ideal gas of rigid molecules. These need to be combined
with good approximations for the excess functional $F\sub{ex}[\{N_\alpha\}]$
to obtain a practicable theory for inhomogeneous liquids.

\subsection{Excess functionals for model fluids}
The fluid of hard spheres has been studied extensively
theoretically as well as with computer simulations.
Within classical density-functional theory, it is
accurately described by Rosenfeld's fundamental measure theory
\cite{RosenfeldFMT}, which satisfies several rigorous conditions
such as reducing to the exact Percus functional in the inhomogeneous
one dimensional limit \cite{HardRodExact-Percus}
and reproducing the Percus-Yevick pair correlations
\cite{PYHardSphere-Wertheim} in the bulk three dimensional limit.

There are several variants of the fundamental measure theory functional
corresponding to different bulk equations of state and regularizations
for the zero-dimensional limit. (See \cite{FMTreview} for a detailed review.)
The excess functional $F\sub{ex}$ for the highly accurate
`White Bear mark II' variant \cite{WhiteBearFMT_markII} based on the
Carnahan-Starling equation of state for the bulk hard sphere fluid \cite{CarnahanStarlingEOS},
including tensor regularizations due to Tarazona \cite{TarazonaFMT}, is
\begin{flalign}
\Phi\sub{HS}[N] &= T \int \textrm{d}\vec{r} \left(
	\begin{aligned}
	& n_0 \log\frac{1}{1-n_3} + f_2(n_3) \frac{n_1n_2
		- \vec{n}_{v1}\cdot\vec{n}_{v2}}{1-n_3} +\\
	& f_3(n_3) \frac{n_2^3 - 3 n_2 |\vec{n}_{v2}|^2
		+ 9\left(\vec{n}_{v2}\cdot\tensor{n}_{m2}\cdot\vec{n}_{v2}
			- \textrm{Tr}\frac{\tensor{n}_{m2}^3}{2}\right)}
		{24\pi(1-n_3)^2}
	\end{aligned}
\right), \label{eqn:FMT} \\
\textrm{with} & \nonumber\\
f_2(n_3) &\equiv 1 + \frac{n_3(2-n_3) + 2(1-n_3)\log(1-n_3)}{3n_3} \textrm{ and}\nonumber\\
f_3(n_3) &\equiv 1 - \frac{2 n_3 - 3 n_3^2 + 2 n_3^3 + 2(1-n_3)^2\log(1-n_3)}{3 n_3^2},\nonumber
\end{flalign}
where the $n_i$'s are scalar ($i=0,1,2,3$), vector ($i=v1,v2$)
and rank-2 tensor ($i=m2$) weighted densities defined as
$n_i(\vec{r}) \equiv w_i \ast N \equiv \int \textrm{d}\vec{r'} w_i(\vec{r}-\vec{r}') N(\vec{r}')$
for hard sphere density $N(\vec{r})$.
The weight functions $w_i$ are spherical measures of various
dimensions (volume, surface etc.) given by
\begin{flalign}
w_0(\vec{r}) &= \delta(R\sub{HS}-r)/(4\pi r^2) \nonumber\\
w_1(\vec{r}) &= \delta(R\sub{HS}-r)/(4\pi r) \nonumber\\
w_2(\vec{r}) &= \delta(R\sub{HS}-r) \nonumber\\
w_3(\vec{r}) &= \theta(R\sub{HS}-r) \nonumber\\
\vec{w}_{v1}(\vec{r}) &= \frac{\vec{r}}{r}\delta(R\sub{HS}-r) \nonumber\\
\vec{w}_{v2}(\vec{r}) &= \frac{\vec{r}}{4\pi r^2}\delta(R\sub{HS}-r) \nonumber\\
\tensor{w}_{m2}(\vec{r}) &= \left(\frac{\vec{r}\vec{r}}{r^2} - \frac{1}{3}\tensor{1} \right)\delta(R\sub{HS}-r)
\label{eqn:FMTweights}
\end{flalign}

The hard sphere fluid also serves as an excellent reference
for perturbation theory for other model systems.
For example, the pair-potential for the Lennard-Jones fluid
\begin{flalign}
U\sub{LJ} &= 4\epsilon \left[\left(\frac{\sigma}{r}\right)^{12}-\left(\frac{\sigma}{r}\right)^6 \right]
\end{flalign}
with energy scale parameter $\epsilon$ and range parameter $\sigma$
is often split into repulsive and attractive parts \cite{LJdecomposition} as
\begin{flalign}
U_R(r) &=
	\begin{cases}
	\epsilon+4\epsilon \left[\left(\frac{\sigma}{r}\right)^{12}-\left(\frac{\sigma}{r}\right)^6 \right], & r<2^{1/6}\sigma \\
	0, & r\ge2^{1/6}\sigma
	\end{cases} \\
U_A(r) &=
	\begin{cases}
	-\epsilon, & r<2^{1/6}\sigma \\
	4\epsilon \left[\left(\frac{\sigma}{r}\right)^{12}-\left(\frac{\sigma}{r}\right)^6 \right], & r\ge2^{1/6}\sigma.
	\end{cases}
\end{flalign}
The free energy functional for this fluid can be
approximated by treating the fluid interacting with
$U_R(r)$ alone using fundamental measure theory,
typically with a hard sphere radius $R\sub{HS} = \sigma/2$,
and then accounting for the effects of $U_A(r)$  perturbatively.
Mean field perturbation then leads to the excess functional
\begin{equation}
F\sub{ex}\super{(MF)}[N(\vec{r})] \approx \Phi\sub{HS}[N]
	+ \frac{1}{2}\int \textrm{d}\vec{r}\int\textrm{d}\vec{r}' N(\vec{r}) U_A(|\vec{r}-\vec{r}'|) N(\vec{r}'),
\end{equation}
and several beyond-mean-field approaches have been developed to improve upon this starting point.

Of particular interest is the recent approach of Peng and Yu \cite{LJ-MWFDFT}
to recast the mean-field term into a nonlinear weighted-density form
\begin{equation}
F\sub{ex}\super{(MWF)}[N(\vec{r})] \approx \Phi\sub{HS}[N]
	+ \int \textrm{d}\vec{r} N(\vec{r}) A\sub{att}\super{LJ}(w_A \ast N), \label{eqn:LJ-MWFDFT}
\end{equation}
with the mean-field weight function set to the normalized perturbation potential
\begin{equation}
w_A(r) = \frac{U_A(r)}{\int 4\pi r'^2 \textrm{d}r' U_A(r')}
	= \frac{9}{8\sqrt{2}\pi\sigma^3}
	\begin{cases}
	1/4, & r<2^{1/6}\sigma \\
	\left(\frac{\sigma}{r}\right)^{6}-\left(\frac{\sigma}{r}\right)^{12}, & r\ge2^{1/6}\sigma.
	\end{cases} \label{eqn:LJ-MWF}
\end{equation}
Here, $A\sub{att}\super{LJ}(N)\equiv A\sub{LJ}(N) - A\sub{HS}(N)$
is the difference between the Helmholtz energy per particle
for the uniform Lennard-Jones fluid and the uniform hard sphere fluid
at the same bulk density $N$. Peng and Yu demonstrate that this functional
does an excellent job of reproducing the inhomogeneous density profiles
and vapor-liquid interface energies in comparison to Monte Carlo
simulations of the Lennard-Jones fluid.

\subsection{Excess functional for liquid water}  \label{sec:ScalarEOS}

The situation for a polar molecular fluid such as water
is much more complicated than the model fluids mentioned above.
Most approaches to the excess free energy of inhomogeneous
water \cite{WDA,WaterFreezingDFT,LischnerH2O} are constructed
to reproduce the pair-correlations in the uniform fluid limit
obtained by computer simulations or from neutron-scattering data.
They can be reasonably accurate for modest inhomogeneities,
but their practicality is limited as they are tied to the
temperature and pressure of the simulation/experiment data
that they are based on, and usually lack a simple analytic formulation.

An alternate strategy is based on identifying a simple model Hamiltonian
for which an approximate analytic free energy functional is readily formulated,
and then constraining the parameters of the model Hamiltonian to 
the bulk properties of the fluid, such as the equation of state.
Wertheim's thermodynamic perturbation theory \cite{WertheimTPT}
is a useful framework for generating free energy functionals;
one class of Hamiltonians considered for water within this framework 
is based on tetrahedral association sites for hydrogen bonds \cite{SAFT-Water},
but these models are yet to successfully predict the quantities relevant to
solvation such as pair correlations, cavitation energies and dielectric response,
partly due to the relative complexity of the model Hamiltonian.

Recently, we proposed an alternate model Hamiltonian \cite{BondedVoids}
based on capturing the effects of the empty space in the tetrahedral hydrogen bond network
by attaching `void' spheres to the molecule in the directions conjugate to
the tetrahedral hydrogen-bond directions. The bonding constraints
in the resulting rigid trimers of hard spheres was also treated using
Wertheim perturbation theory, but the relative simplicity of that model
enabled an accurate free energy functional description of the inhomogeneous fluid
capable of predicting the aforementioned quantities relevant for solvation.

This `bonded-voids' free energy functional for water is adequately
accurate for cavitation energies, dielectric response and the height
and particle content of the first peak in the pair correlation.
However, the secondary peaks in its pair correlation occur at the
characteristic distances for a close-packed hard sphere fluid
rather than for a tetrahedrally-bonded one.
Evidently the cavitation energies are not sensitive to this
deficiency in the secondary structure of the pair correlation;
the height of the first peak and the exclusion volume (location of pole)
in the equation of state are the important factors,
which are captured correctly by the bonded void spheres ansatz.

Here, we present a simplified free energy functional for water
which retains only the critical features of the bonded-voids model
\cite{BondedVoids}, while eliminating the complexity of Wertheim perturbation.
This functional employs a hard sphere reference with a weighted density term
constrained to reproduce the equation of state in the spirit
of the approach of \cite{LJ-MWFDFT} for the Lennard-Jones fluid.
Due to the polar nature, we need to distinguish between
short-ranged orientation-averaged interactions with a $r^{-6}$
tail similar to the Lennard-Jones pair potential
and long-range orientation-dependent interactions with a $r^{-1}$
tail between individual charged sites resulting in $r^{-3}$
for neutral molecules with a net dipole moment.

We deal with the long range orientation-dependent part
by taking advantage of the rigid molecule site-model
capability developed in section \ref{sec:IdealGas}.
In particular, we adopt the molecule geometry and site charges
of the popular SPC/E pair potential model \cite{SPCE} for
molecular dynamics simulations of water, which consists of
an $O$ site with charge $Z_O=+0.8476~e^-$ and two $H$ sites with
charge $Z_H=-0.4238~e^-$ in a bent geometry with an
$O$-$H$ distance of $1$~\AA~and a tetrahedral $H$-$O$-$H$
angle ($\cos^{-1}(-1/3)\approx109.5^\circ$).

For the shorter-ranged orientation-dependent part,
we assume a Lennard-Jones interaction between the
$O$-sites since it has the correct $r^{-6}$ tail.
We arrive at the excess functional ansatz
\begin{multline}
F\sub{ex}^{H_2O}[N_O(\vec{r}),N_H(\vec{r})] 
	\approx \Phi\sub{HS}[N_O]
		+ \int \textrm{d}\vec{r} N_O(\vec{r}) A\sub{att}^{H_2O}(w_A \ast N_O) \\
	+ \frac{A_\epsilon(T)}{2} \sum_{\alpha,\beta\in\{O,H\}} Z_\alpha Z_\beta
		\int\textrm{d}\vec{r}\int\textrm{d}\vec{r}' N_\alpha(\vec{r}) K(|\vec{r}-\vec{r}'|) N_\beta(\vec{r}'),
	\label{eqn:Fex-ScalarEOS}
\end{multline}
by adding a long-range polar correction (third term)
to the Lennard-Jones functional of \cite{LJ-MWFDFT} (first two terms).
The following paragraphs specify the Helmholtz energy function $A\sub{att}^{H_2O}(N)$,
the dipole correlation factor $A_\epsilon(T)$ and the modified Coulomb kernel $K(r)$.
We shall refer to this excess functional (\ref{eqn:Fex-ScalarEOS})
as `scalar-EOS' because the excess free energy density
is attributed to the scalar moment of the orientation density
and is constrained to the equation of state.

In (\ref{eqn:Fex-ScalarEOS}), $\Phi\sub{HS}$ is the White Bear mark II
fundamental theory functional, given by (\ref{eqn:FMT}),
for a fluid of hard spheres of radius $R\sub{HS}$.
The second weighted density term employs the mean-field weight
function $w_A(r)$ given by (\ref{eqn:LJ-MWF})
with $\sigma=2R\sub{HS}$.

The third term of (\ref{eqn:Fex-ScalarEOS}) is the mean-field
electrostatic interaction between the charge-site densities
scaled by a dipole-correlation factor $A_\epsilon(T)$.
Following \cite{LischnerH2O}, the Coulomb kernel $K(r)$
is cutoff at high frequencies as
\begin{equation}
\tilde{K}(G) = \frac{4\pi}{G^2} \left[1 + \left(\frac{G}{G_c}\right)^4\right]^{-1} \label{eqn:BWlimitedCoulomb}
\end{equation}
with $G_c = 0.33$ bohr$^{-1}$,
and the dipole correlation factor is chosen to
reproduce the bulk linear dielectric constant.
Without the correlation factor, i.e. with $A_\epsilon=1$,
the SPC/E geometry would yield a dielectric constant of $19.7$
at ambient conditions instead of the experimental value of $78.4$.
The single parameter fit
\begin{equation}
A_\epsilon(T) = 1 - \frac{T}{7.35\times10^3~\textrm{K}}
\end{equation}
reproduces the bulk linear dielectric constant over the 
entire liquid phase with a relative RMS error $\sim 1\%$.

Next, we constrain $F\sub{ex}^{H_2O}$ to reproduce
the correct Helmholtz energy density for the uniform fluid
of molecular density $N$, which may be obtained by
integrating the equation of state ($p(N,T)$).
Note that the third term of (\ref{eqn:Fex-ScalarEOS}) does not
contribute to the uniform fluid free energy, and hence 
$A\sub{att}^{H_2O}$ must be the difference between the per-molecule
Helmholtz free energy in water and the hard sphere fluid.
Using the Jefferey-Austin equation of state \cite{JeffAustinEOS}
for water, this constrains
\begin{multline}
A\sub{att}^{H_2O}(N) =
	\frac{\alpha T}{\lambda b(T)} \log\frac{1}{1-\lambda b(T) N} - (a\sub{VW}+b^\ast T) N \\
	- 2 T f^{\ast\ast}(T) \frac{1+C_1}{1+C_1 \exp\frac{(N-\rho\sub{HB})^2}{\sigma^2}}
		\log \frac{\Omega_0+\Omega\sub{HB}e^{-\epsilon\sub{HB}/T}}{\Omega_0+\Omega\sub{HB}} \\
	- T \frac{V\sub{HS}N(4-3V\sub{HS}N)}{(1-V\sub{HS}N)^2}
\label{eqn:H2O-Aatt}
\end{multline}
up to a temperature-dependent constant which is absorbed
into the arbitrary reference for the chemical potential $\mu$.
The first two lines of (\ref{eqn:H2O-Aatt}) represent the
free energy density corresponding to the excess pressure
for liquid water as parametrized in \cite{JeffAustinEOS}
by fits to experimental data for bulk liquid water,
and the definitions of the numerous constants and
functions of temperature may be found therein.\footnote{
Note that the constants listed in \cite{JeffAustinEOS}
are in SI/CGS units, and should be converted to atomic units
(with $k_B=1$) before substitution in (\ref{eqn:H2O-Aatt}).}
The last line of (\ref{eqn:H2O-Aatt}) subtracts
the uniform fluid per-particle free energy
corresponding to $\Phi\sub{HS}$ given by (\ref{eqn:FMT}),
with $V\sub{HS}=4\pi R\sub{HS}^3/3$.

Now, (\ref{eqn:Fex-ScalarEOS}) is completely specified except for
the value of the hard sphere radius $R\sub{HS}$.
Unlike the Lennard-Jones fluid, there is no prescribed
pair potential from which it may be derived.
We require that calculations with the excess functional (\ref{eqn:Fex-ScalarEOS})
result in the surface-energy of the planar water liquid-vapor interface in agreement
with the experimental surface tension of \sci{72.0}{-3}~N/m
at ambient temperature 298~K, and obtain
\begin{equation}
R\sub{HS} =  1.36~\textrm{\AA}
\end{equation}
The details of the planar interface calculation
are presented in Section~\ref{sec:Discretization},
and tests of the accuracy of the scalar-EOS functional for
inhomogeneous liquid water are in Section~\ref{sec:WaterAccuracy}.

\section{Results} \label{sec:Results}

The efficient rigid-molecular ideal gas representations of
section \ref{sec:IdealGas} combined with the
excess functional for water from section \ref{sec:ScalarEOS}
forms a practical theory of inhomogeneous liquid water as we show below.
We use this system to study the convergence properties of
the various molecular ideal gas representations in section \ref{sec:Convergence},
and then test the accuracy of the  scalar-EOS water functional 
against experiment and molecular dynamics simulations
in section \ref{sec:WaterAccuracy}.

\subsection{Discretization} \label{sec:Discretization}

The free energy functional approximations presented here
involve integrals over space and orientations, which must
all be discretized in a practical calculation.
The discretization of three dimensional space may be performed in a variety of bases
including plane-waves, wavelets and specialized bases such
as planar and radial one dimensional grids for high symmetry cases.

We present the details of the numerical formulation
of the free energy functionals for rigid-molecular liquids
using the basis-independent algebraic formulation
developed for electronic density-functional theory \cite{AlgebraicDFT}.
Within this formulation, the physics is expressed
in terms of a handful of abstract operators
independent of the basis, while the implementation
of these operators in code is basis dependent.
This allows for the same top-level physics code
to be used with multiple basis sets with no modification.
A three-dimensional plane-wave basis implementation of the
fluid framework and excess functionals (using the notation
and operators described below) is distributed with the open-source
electronic density-functional theory software JDFTx \cite{JDFTx},
which specializes in solvated \emph{ab initio} calculations.
An analogous code base for high-symmetry one-dimensional basis sets,
suitable for development and testing of new fluid functionals,
is distributed as a sub-project of JDFTx \cite{Fluid1D}.

Here, we briefly introduce the notation and operators
required for classical density-functional theory;
see \cite{AlgebraicDFT} for a detailed description.
A function of space $f(\vec{r})$ is expanded in terms
of basis functions $\{b_i(\vec{r})\}$ with coefficients
$\tilde{f}_i$ (often written as a vector $\tilde{f}$)
i.e. $f(\vec{r}) = \sum_i \tilde{f}_i b_i(\vec{r})$.

The overlap of two functions $f(\vec{r})$ and $g(\vec{r})$ is
\begin{equation}
\int \textrm{d}\vec{r} f^\ast(\vec{r}) g(\vec{r})
= \sum_{i,j} \tilde{f}_i^\ast \tilde{g}_j 
	\underbrace{\int \textrm{d}\vec{r} b_i^\ast(\vec{r}) b_j(\vec{r})}_{\mathcal{O}_{ij}}
= \tilde{f}^\dag \mathcal{O} \tilde{g}
\end{equation}
which defines the basis overlap matrix $\mathcal{O}$
(which would be diagonal for orthogonal basis sets).
Similarly, any linear operator reduces to a matrix. For example,
$\int \textrm{d}\vec{r} f^\ast(\vec{r}) \nabla^2 g(\vec{r}) = \tilde{f}^\dag \mathcal{L} \tilde{g}$
defines the Laplacian matrix $\mathcal{L}_{ij} = \int \textrm{d}\vec{r} b_i^\ast(\vec{r})\nabla^2 b_j(\vec{r})$.

The density functionals also involve integrals over nonlinear functions
which of course cannot be reduced to basis-space matrices
like the linear operators considered above.
Consequently, the basis sets are accompanied by a quadrature grid
consisting of a set of nodes $\{\vec{r}_j\}$
over which integration of nonlinear functions is performed.
A function $f(\vec{r})$ sampled on this quadrature grid
$f_j = f(\vec{r}_j)$ is denoted simply by the vector $f$.
This introduces the linear basis-to-real space operator
$\mathcal{I}$ defined by $f = \mathcal{I} \tilde{f}$
with matrix elements $\mathcal{I}_{ji} = b_i(\vec{r}_j)$, and the real-to-basis
 space operator, $\mathcal{J} = \mathcal{I}^{-1}\sub{left}$.\footnote{
$\mathcal{J} = \mathcal{I}^{-1}$ is the natural choice when the
number of basis functions equals the number of quadrature grid points,
which is the case for the plane-wave basis for example.
When the number of grid points exceeds the number of basis functions,
one possibility is to use the left-inverse as indicated
so that $\mathcal{JI}=1$, although this is not necessary.}
Armed with these operators, we can discretize the commonly encountered integral
$\int\textrm{d}\vec{r}f(\vec{r}) A(g(\vec{r}))
= \tilde{f}^\dag\mathcal{OJ} A(\mathcal{I}\tilde{g})
= f^\dag\mathcal{J^\dag OJ} A(g)$
where $A$ is some nonlinear function (which operates element-wise on vectors).

In the particular case of plane-wave basis on a periodic unit cell,
the quadrature grid $\vec{r}_j$ is a uniform parallelepiped mesh,
the basis functions are $e^{-i\vec{G}\cdot\vec{r}}$
for reciprocal lattice vectors $\vec{G}$,
and the operators $\mathcal{I}$ and $\mathcal{J}$
are implemented as Fast Fourier Transforms.
$\mathcal{O}$ is the scalar matrix $\Omega$,
and $\mathcal{L}$ is the diagonal matrix $-\Omega|\vec{G}|^2$,
where $\Omega$ is the unit cell volume. For a detailed specification of these operators,
see \cite{AlgebraicDFT} for the three-dimensional plane-wave basis,
\cite{AriasWavelets} for a multi-resolution (wavelet) basis,
and \ref{sec:Discretization1D} for the planar,
cylindrical and spherical one-dimensional grids.

In fact, the six operators introduced above 
(counting hermitian adjoints separately) are the
\emph{only} ones required for electronic density
functional theory in the local density approximation (LDA).
The advantage of writing code in this framework is that
implementing a new basis only requires
reimplementing the small number of these operators.

To express the classical density functionals,
we need to introduce two additional operators.
Firstly, the computation of weighted densities involves convolutions
$h(\vec{r}) = \int d\vec{r}' f(\vec{r}-\vec{r}') g(\vec{r}')$,
which may be discretized using a basis dependent tensor $\mathcal{C}^{k}_{ij}$
to $\tilde{h}_k = \sum_{i,j} \mathcal{C}^{k}_{ij} \tilde{f}_i \tilde{g}_j$,
which we also denote by $\tilde{h} = \tilde{f} \ast \tilde{g}$ for brevity.
Integrating the defining relation multiplied by basis functions,
we see that the convolution tensor elements must be
\begin{equation}
\mathcal{C}^k_{ij} = \sum_l (\mathcal{O}^{-1})_{kl}
	\int\textrm{d}\vec{r}\int\textrm{d}\vec{r}'
		b_l^\ast(\vec{r}) b_i(\vec{r}-\vec{r}') b_j(\vec{r}').
\end{equation}
$\mathcal{C}^k_{ij}$ is symmetric under $i\leftrightarrow j$
when the space is translationally invariant,
and reduces to the element-wise multiply 
$\mathcal{C}^k_{ij} = \Omega \delta_{ki}\delta_{kj}$
for the plane-wave basis, as is well known.

Secondly, the rigid molecule formalism of section \ref{sec:IdealGas}
requires sampling functions with arbitrary displacements in order to
generate orientation densities from the effective site potentials,
and to generate the site densities from the orientation densities.
This requires the inclusion of a translation operator defined by 
$\mathcal{T}_{\vec{a}} f(\vec{r}) = f(\vec{r}+\vec{a})$ to our toolkit.
This may be discretized to $\tilde{g}_i = \sum_j (\mathcal{T}_{\vec{a}})_{ij}\tilde{f}_j$
where $\tilde{f}$ and $\tilde{g}$ are the discretizations of
$f(\vec{r})$ and $f(\vec{r}+\vec{a})$ respectively.
The natural translation operator for a given basis set obtained
by integrating the definition multiplied by basis functions is
\begin{equation}
(\mathcal{T}_{\vec{a}})_{ij} = \sum_k (\mathcal{O}^{-1})_{ik}
	\int\textrm{d}\vec{r} b_k^\ast(\vec{r}) b_j(\vec{r}+\vec{a}),
\end{equation}
and satisfies $\mathcal{T}_{\vec{a}}^\dag = \mathcal{T}_{-\vec{a}}$
by definition. In the plane-wave basis, this operator takes the diagonal form
$(\mathcal{T}_{\vec{a}})_{ij} = \delta_{ij}e^{-i\vec{G}_i\cdot\vec{a}}$.

However, this `Fourier' translation operator
introduces severe ringing in functions that have
components that extend up to the Nyquist frequency.
This can be quite problematic for the classical
density-functional theory of rigid molecules,
particularly since positive functions can ring
negative on translation, leading to regions of
negative site densities even when $p_\omega\ge0$.
The free energy functionals evaluated for negative
site densities can be unphysically favorable
which encourages further ringing,
resulting in a numerical divergence.\footnote{
In principle, the contributions to the free energy
from regions of negative site densities could be redefined to zero.
However, this results in a highly non-analytic energy landscape
with extremely poor convergence for minimization algorithms}

We remedy this by using inexact translation operators which
have the property that they map the set of functions with
all-non-negative samples on the quadrature grid onto itself.
The action of the translation operator on the quadrature grid
$\mathcal{S}_{\vec{a}} \equiv \mathcal{IT}_{\vec{a}}\mathcal{J}$
can be viewed as sampling the function on the grid with
displacement $\vec{a}$. The natural translation operator for
the plane-wave basis corresponds to a sampling operator
$\mathcal{S}$ based on Fourier interpolation.
Amongst the piece-wise polynomial spline interpolations,
only the constant spline (pick nearest neighbor)
and linear spline (linear interpolation in each cell)
guarantee non-negative results for a non-negative sample set;
we denote the corresponding approximate sampling operators
by $\mathcal{S}^{C}$ and $\mathcal{S}^{L}$ respectively.

The discretization of spatial integrals in the rigid-molecule classical
density functional framework can be achieved using the above operators;
the final ingredient is the discretization of the orientation integrals.
We achieve this using a quadrature rule directly on $SO(3)/G$,
where $G$ is the symmetry group of the fluid molecule, so that
\begin{equation}
\int_{\omega\in SO(3)} \frac{\textrm{d}\omega}{8\pi^2} f(\omega) \to \sum_i W_i f(\omega_i)
\end{equation}
with a finite set of orientations $\omega_i$ and weights $W_i$.
\ref{sec:SO3quad} describes various methods for the generation
of quadrature rules on $SO(3)/\mathbb{Z}_n$ ranging from
outer product quadratures on Euler angles to uniform sampling
sets based on platonic solid rotation groups. Section~\ref{sec:Convergence}
explores the convergence of the orientation integrals with quadrature
for the scalar-EOS water functional (symmetry group $\mathbb{Z}_2$),
and the list of explored quadratures is summarized in Table~\ref{tab:SO3quad}.

We can now discretize the molecular ideal gas free energy (\ref{eqn:PhiIDfromPomega})
given the orientation density $p_{\omega_i}$ on the quadrature grid
for each discrete orientation and the site densities $\tilde{N}_\alpha$
in the chosen basis set, as
\begin{equation}
\Phi\sub{id} =
	T \tilde{1}^\dag\mathcal{OJ}\sum_i W_i p_{\omega_i} \left(\log\frac{p_{\omega_i}}{N_\textrm{ref}} - 1\right)
	+ \sum_\alpha \tilde{N}_\alpha^\dag \mathcal{O} (\tilde{V}_\alpha - \mu_\alpha\tilde{1})
	\label{eqn:DiscretePhiIDfromPomega}
\end{equation}
Note that all unary real functions are understood to operate element-wise
on vectors on the quadrature grid, unless otherwise specified.

The expression of the orientation density on the quadrature grid in terms of
the independent variables for minimization depends on the chosen representation.
In the self representation, the independent variables are $\tilde{p}_{\omega_i}$
in basis space and therefore $p_{\omega_i} = \mathcal{I} \tilde{p}_{\omega_i}$.
The independent variables in the site-potential representation are $\tilde{\psi}_\alpha$
and the orientation density is generated using (\ref{eqn:PomegaFromPsiAlpha}) as
\begin{flalign}
p_{\omega_i} &= N\sub{ref} \exp \left(\frac{-1}{T} \mathcal{I}\sum_{\alpha,k} 
	\mathcal{T}_{\omega_i\circ\vec{R}_{\alpha k}}\tilde{\psi}_\alpha \right)\nonumber\\
	&= N\sub{ref} \exp \left(\frac{-1}{T} \sum_{\alpha,k} 
	\mathcal{S}_{\omega_i\circ\vec{R}_{\alpha k}} \mathcal{I}\tilde{\psi}_\alpha \right),
	\label{eqn:DiscretePomegaFromPsiAlpha}
\end{flalign}
where the latter expression with an approximate sampling operator $\mathcal{S}$
is used in practice. In the multipole representation, the independent variables
are $\tilde{\mu}^j_{m_1m_2}(\vec{r})$ for $|m_1|,|m_2| \le j \le j\sub{max}$
and the orientation density is generated using (\ref{eqn:PomegaFromMujmm}) as
\begin{equation}
p_{\omega_i} = N\sub{ref} \exp \left(\frac{-1}{T} \mathcal{I} \sum_{j=0}^{j\sub{max}} \sum^{+j}_{m_1,m_2=-j}
	D^j_{m_1 m_2}(\omega_i) \tilde{\mu}^j_{m_1 m_2}  \right),
	\label{eqn:DiscretePomegaFromMujmm}
\end{equation}
which simplifies for $j\sub{max}=1$ in terms of independent variables
$\tilde{\mu}$ and $\vec{\tilde{\epsilon}}$ to
\begin{equation}
p_{\omega_i} = N\sub{ref} \exp \frac{-\mathcal{I} \left( \tilde{\mu}
		+ (\omega_i\circ\hat{z})\cdot\vec{\tilde{\epsilon}} \right) }{T}.
	\label{eqn:DiscretePomegaFromMuEps}
\end{equation}

Finally, the site densities are generated from the orientation density
by a discretization of (\ref{eqn:NalphaFromOmegaNI}), given by
$N_\alpha \equiv \frac{\delta}{\delta\psi_\alpha}\Omega^{(ni)}$,
with $\Omega^{(ni)} = -T \tilde{1}^\dag\mathcal{OJ}\sum_i W_i p_{\omega_i}$
and $p_{\omega_i}$ given by (\ref{eqn:DiscretePomegaFromPsiAlpha}),
so that\footnote{This is derived from $d\Omega =
\int \frac{\delta\Omega}{\delta\psi} d\psi = \tilde{1}^\dag\mathcal{OJ}
\textrm{Diag}(\frac{\delta\Omega}{\delta\psi}) d\psi$, which leads to
$\textrm{Diag}(\mathcal{J}^\dag\mathcal{O}\tilde{1})
\frac{\delta\Omega}{\delta\psi} = \frac{\partial\Omega}{\partial\psi^\dag}$.
Here, $\textrm{Diag}(x)$ is the diagonal operator with the elements of $x$
on its diagonal, i.e. $\textrm{Diag}(x)y = \textrm{Diag}(y)x$
is the element-wise multiplication of $x$ and $y$.}
\begin{equation}
N_\alpha = \textrm{Diag}(\mathcal{J}^\dag\mathcal{O}\tilde{1})^{-1}
	\sum_i W_i \sum_k  \mathcal{S}^\dag_{\omega_i\circ\vec{R}_{\alpha k}}
	\textrm{Diag}(\mathcal{J}^\dag\mathcal{O}\tilde{1}) p_{\omega_i}.
	\label{eqn:DiscreteNalphaFromPomega}
\end{equation}
For the translationally invariant plane-wave basis set,
the above expression is equivalent to $N_\alpha = \sum_i W_i
\sum_k \mathcal{S}_{-\omega_i\circ\vec{R}_{\alpha k}} p_{\omega_i}$,
the intuitive discretization of (\ref{eqn:NalphaFromPomega}),
and this holds approximately for other three-dimensional basis sets.
However, (\ref{eqn:DiscreteNalphaFromPomega}) holds even when
$\mathcal{S}_{\vec{a}}$ is generalized to a non-uniform translation
$\mathcal{S}_{\vec{a}(\vec{r})}$, which is required for the
reduced-dimensionality basis sets of \ref{sec:Discretization1D}.

Moving on to excess functionals, the hard sphere excess free energy
$\Phi\sub{HS}[N]$ given by (\ref{eqn:FMT}) is discretized by replacing
$\int\textrm{d}\vec{r} \to \tilde{1}^\dag\mathcal{OJ}$
and computing the integrand element-wise on the quadrature grid,
where the weighted densities are computed from convolutions
in basis space $n_i = \mathcal{I}(\tilde{w}_i \ast \tilde{N})$.
These convolutions may be computed efficiently in the plane-wave basis by multiplying
with the analytic Fourier transforms of the weight functions (\ref{eqn:FMTweights}),
but in other bases, they should be computed with specialized routines that take
advantage of the finite range of the weight functions. (See \cite{FMTreview} for examples.)
The excess free energy of the Lennard-Jones fluid \cite{LJ-MWFDFT},
given by (\ref{eqn:LJ-MWFDFT}), discretizes to
$F\sub{ex}\super{(MWF)} = \Phi\sub{HS}[N] + \tilde{N}^\dag\mathcal{OJ}
	A\sub{att}\super{LJ}(\mathcal{I}(\tilde{w}_A \ast \tilde{N}))$.
The convolution $\tilde{w}_A \ast$ is trivial in the plane-wave basis,
but may require specialized routines in other basis sets due to
the polynomial tail of the Lennard-Jones weight function.\footnote{
For example, in wavelet bases, this may be performed by decomposition
into a finite-ranged part treated at all grid levels, and a bandwidth-limited
long-range part performed using the Fourier method on the coarsest grid.}

Finally, the scalar-EOS excess functional for water (\ref{eqn:Fex-ScalarEOS})
is discretized to
\begin{multline}
F\sub{ex}^{H_2O} = \Phi\sub{HS}[N_O]
	+ \tilde{N}_O^\dag\mathcal{OJ} A\sub{att}^{H_2O}(\mathcal{I}(\tilde{w}_A \ast \tilde{N}_O)) \\
	+ \frac{A_\epsilon(T)}{2} \sum_{\alpha,\beta\in\{O,H\}} Z_\alpha Z_\beta
		(\tilde{w}_K\ast\tilde{N}_\alpha)^\dag 
		\bar{\mathcal{O}} (-4\pi\mathcal{L}^{-1}) \bar{\mathcal{O}}
		(\tilde{w}_K\ast\tilde{N}_\beta).
	\label{eqn:DiscreteFex-ScalarEOS}
\end{multline}
Here, the high-frequency cutoff Coulomb Kernel (\ref{eqn:BWlimitedCoulomb})
has been rewritten in terms of the bare Coulomb kernel
$(-4\pi\mathcal{L}^{-1}) \bar{\mathcal{O}}$
computed by solving the Poisson equation\footnote{
$\bar{\mathcal{O}}$ is the overlap operator with the null-space of $\mathcal{L}$
projected out, and $\mathcal{L}^{-1}$ is understood to be the inverse of $\mathcal{L}$
in orthogonal complement of the null-space with zero projection in the null-space.
See \cite{AlgebraicDFT} for details.},
by introducing the site-charge kernel
\begin{equation}
\tilde{w}_K(G) = 1/\sqrt{1 + \left(\frac{G}{G_c}\right)^4}. \label{eqn:SiteChargeKernel}
\end{equation}
This modification has no effect for the plane-wave basis, but is important
for other basis sets because it decomposes the long-range convolution with $\tilde{K}(G)$
into a short-ranged convolution ((\ref{eqn:SiteChargeKernel}) is confined
exponentially in real space), and the solution of Poisson equation
which is a standard operation in any basis set \cite{AlgebraicDFT,AriasWavelets}.

\subsection{Convergence} \label{sec:Convergence}

Section~\ref{sec:Discretization} presented the discretization of the
general rigid-molecular ideal gas framework of section~\ref{sec:IdealGas}
with various choices for the independent variables, and excess functionals
including the scalar-EOS functional for liquid water constructed in
section~\ref{sec:ScalarEOS}. Next, we briefly discuss the minimization
of the liquid free energy given a set of external potentials,
compare the performance of the different choices of independent variables,
and explore the accuracy of the discretization of the orientation integrals.

The free energy of the fluid for a particular excess functional
and choice of independent variables is expressed in the basis-independent
algebraic formulation of \cite{AlgebraicDFT}, including the operators
introduced in section~\ref{sec:Discretization}. The gradient of the
free energy with respect to the independent variables may be derived
in the same notation in a straightforward manner as shown in \cite{AlgebraicDFT},
and the computational cost for evaluating the gradient is comparable
to that for the free energy. We can therefore minimize the free energy
functional to find the equilibrium configuration of the fluid directly
using the non-linear conjugate gradients method \cite{FletcherReevesCG}.

First, we compare the convergence of the conjugate gradients method
for different choices of independent variables. For the remainder
of this section, we work with the scalar-EOS functional for water
at a temperature of 298~K in the three-dimensional plane-wave basis set,
and perform all calculations using JDFTx \cite{JDFTx}.
We focus on two physical systems which capture different extremes
of the typical external potentials encountered in ab-initio solvation:
water surrounding a hard sphere, and water in a parallel plate capacitor
with a strong electric field (in the dielectric saturation limit).

\begin{figure}
\begin{tabular}{cc}
\includegraphics[width=2.3in]{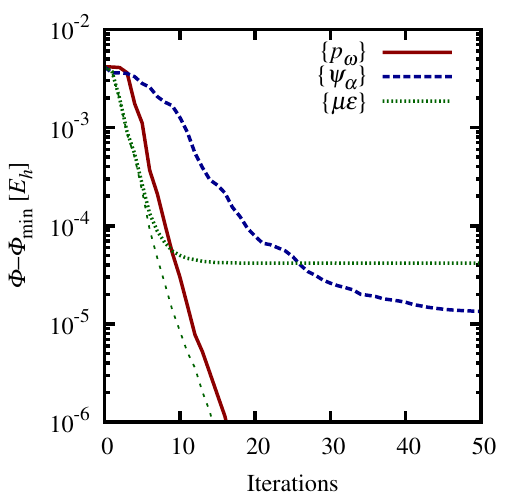}
& \includegraphics[width=2.3in]{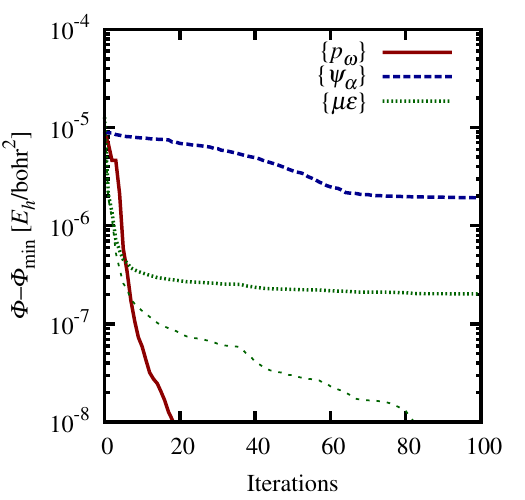} \\
(a) 4~\AA~ hard sphere & (b) Capacitor with $E_0=1$~V/\AA
\end{tabular}
\caption{Convergence of conjugate gradients minimization of the free energy
of scalar-EOS water (a) surrounding a 4~\AA~ hard sphere, and
(b) in a parallel plate capacitor with externally applied field strength $E_0=1$~V/\AA,
for different independent variables. The difference of free energy from the
final equilibrium value as a function of iteration count is shown on a
logarithmic scale for the self representation $\{p_\omega\}$ (solid red line),
the site-potential representation $\{\psi_\alpha\}$ (blue dashed line),
and the multipole representation $\{\mu,\vec{\epsilon}\}$ (thicker green dotted line).
The fainter green dotted line is the difference of the free energy in the
$\{\mu,\vec{\epsilon}\}$ representation from the converged value within that
representation (which is variationally higher than the equilibrium value).
Note the rapid exponential convergence in the self and multipole representations,
compared to the site potential representation.
\label{fig:ConvergenceCG}}
\end{figure}

The hard sphere system consists of an external potential $V_O(\vec{r})
= V_0\theta(R-|\vec{r}|)$ which excludes the $O$ sites of water
from a sphere of radius $R$, with no potential on the $H$ sites
($V_H(\vec{r})=0$). We pick $R = 4$~\AA, a reasonable size for
the region excluded by a molecule solvated in water, and
$V_0 = 1~E_h$ ($\approx$ 27.2~eV) which is sufficient to completely
exclude the liquid from that region. The calculations are performed
in a cubic unit cell of side 32 bohrs ($\approx$ 17~\AA) with a
$128\times128\times128$ fast Fourier transform (FFT) grid;
the grid spacing of 0.25 bohrs corresponds roughly to the
charge density grid of a typical electronic density-functional theory
calculation at a wave-function kinetic energy cutoff of 20~$E_h$.

The parallel plate capacitor system consists of two plates 112 bohrs
apart, with an external potential corresponding to an applied electric
field of $E_0 = 1$~V/\AA~($10^{10}$~V/m), which is typical in the first
solvation shell of a polar molecule, and corresponds to a regime of
strongly non-linear dielectric response. (See Figure~\ref{fig:nonlineareps}.)
Repulsive potentials of strength 1~Eh on both the $O$ and $H$ sites
confine the fluid to the region between the capacitor plates.
The calculation is performed in a periodic cell of length 256 bohrs
containing two capacitors back-to-back so that the cell has no net dipole,
and is sampled using a one-dimensional FFT grid with 4096 points.
The transverse dimensions are translationally invariant, and the
free energies reported are per bohr$^2$ transverse area.

Figure~\ref{fig:ConvergenceCG} shows the convergence of the
Polak-Ribiere variant of the nonlinear conjugate gradients algorithm
\cite{PolakRibiereCG} for the hard-sphere and capacitor systems with the
site-potential ($\{\psi_\alpha\}$), $j=1$ truncated multipole
($\{\mu,\vec{\epsilon}\}$) and self ($\{p_\omega\}$) representations
of the orientation density as independent variables.
The initial guess in each case corresponds to a uniform bulk density
of water in the allowed regions and no density in the disallowed regions,
with a uniform orientation distribution for the sphere geometry,
and a dipolar orientation distribution corresponding to
bulk linear dielectric response for the capacitor geometry.
The 7-design quadrature with 96 nodes on $SO(3)/\mathbb{Z}_2$
(see Table~\ref{tab:SO3quad}) was used for orientation sampling.

The self representation ($\{p_\omega\}$) exhibits the best
exponential convergence, and is the method of choice 
when it is practical to store the orientation density.
The multipole representation ($\{\mu,\vec{\epsilon}\}$)
also converges quite rapidly, but it is a variational approximation
and will result in a higher free energy than that in $\{p_\omega\}$.
Note that for a typical molecule cavity formation (the hard sphere case),
the error in the $\{\mu,\vec{\epsilon}\}$-representation
is $\sim4\times10^{-5}~E_h$ or $\sim0.03$~kcal/mol,
which is negligible in the computation of solvation energies.
Likewise, the relative error in the free energy of the strong-field
capacitor corresponds to an error of less than $\sim1\%$ in
the effective dielectric constant, which again is acceptable
in the calculation of solvation energies. Finally, the site-potential
representation ($\{\psi_\alpha\}$) of \cite{RISM2,LischnerHCl} exhibits
the poorest convergence, particularly in the strong electric field case.
Although the $\{\psi_\alpha\}$ entropy will eventually converge to the
same value as that of the $\{p_\omega\}$ representation, the approximate $\{\mu,\vec{\epsilon}\}$
representation yields a more accurate free energy at practical iteration counts.

\begin{figure}
\begin{tabular}{cc}
\includegraphics[width=2.3in]{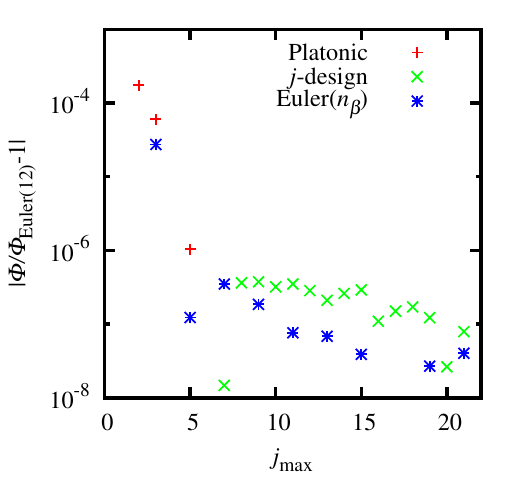}
& \includegraphics[width=2.3in]{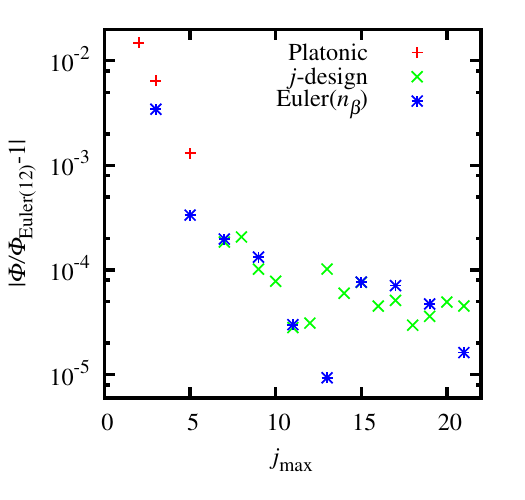} \\
(a) 4~\AA~ hard sphere & (b) Capacitor with $E_0=1$~V/\AA
\end{tabular}
\caption{Convergence of free energy with orientation quadrature for
the two systems considered in Figure~\ref{fig:ConvergenceCG}.
The orientation quadratures studied are listed in Table~\ref{tab:SO3quad},
and the free energy at the Euler(12) quadrature (which has 3456 nodes
on $SO(3)/\mathbb{Z}_2$) is used as the reference in computing
relative errors for all the smaller quadratures. Note that the error due to
the orientation quadrature plateaus at $j_{max}\sim7$ for the sphere geometry,
and at $j_{max}\sim10$ for the strong-field capacitor; these would therefore
be reasonable choices in \emph{ab initio} solvation calculations for
non-polar and strongly-polar molecules respectively.
\label{fig:ConvergenceSO3}}
\end{figure}

\begin{figure}
\begin{tabular}{cc}
\includegraphics[width=1.64in]{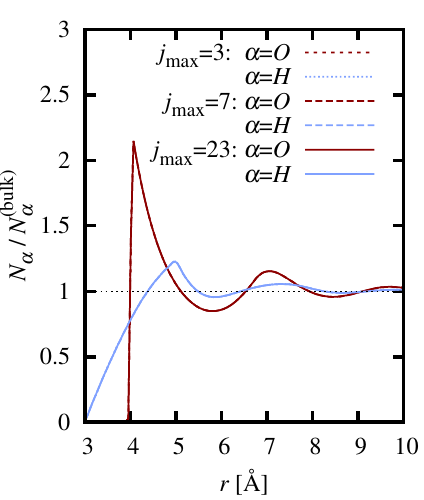}
& \includegraphics[width=2.96in]{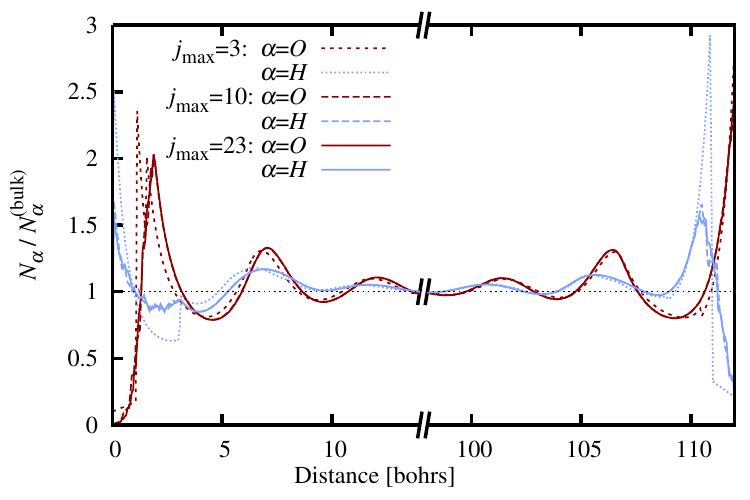} \\
(a) 4~\AA~ hard sphere & (b) Capacitor with $E_0=1$~V/\AA
\end{tabular}
\caption{Convergence of density profiles with orientation quadrature for
the two systems considered in Figure~\ref{fig:ConvergenceCG}.
The plotted site-densities are scaled by their corresponding bulk values,
so that the profiles equilibrate at 1 far from the sphere / plates.
Note that the densities at the lowest and highest quadratures
are indistinguishable for the hard sphere, whereas the densities become
similar to the fully-converged ones only around $j_{max}\sim10$ for
the strong-field capacitor.
\label{fig:DensityConvSO3}}
\end{figure}

Next, we turn to the convergence of the free energies with
respect to the discretization of the orientation integrals.
Figure~\ref{fig:ConvergenceSO3} shows the relative error in the free energy
for each orientation quadrature in Table~\ref{tab:SO3quad} compared to
the Euler(12) quadrature (taken to be the converged value) for the
two systems considered above. The quadratures are sorted by $j_{max}$,
the maximum degree of Wigner functions $D^j_{m_1m_2}$ for which
they are exact. (See \ref{sec:SO3quad} for details.)
The relative error in the free energy decreases rapidly with
quadrature size and plateaus $\sim 10^{-7}$ at $j_{max}\sim7$
for the hard sphere, limited by other discretization errors.
For the highly polarized capacitor, higher quadratures are needed
for the same level of accuracy, and the plateau occurs $\sim 10^{-4}$
at $j_{max}\sim10$. A reasonable choice for $j_{max}$ for a
typical system should therefore range from 7 to 10
depending on the strength of electric fields involved.

Figure~\ref{fig:DensityConvSO3} shows the density profiles next
to the hard sphere and the walls of the capacitor for various
orientation quadratures. In the case of the hard sphere, the density
profiles are virtually identical for all considered quadratures,
as is expected given that the relative error in the free energy
is $\sim10^{-4}$ even for the Octahedron group, one of the
lowest quadratures considered with $j_{max}=3$.
However, there are qualitative differences in the density profiles
near the capacitor walls for $j_{max}=3$ from the converged ones
at $j_{max}=23$ (Euler(12) quadrature), and the differences begin to
disappear only around $j_{max}=10$. At these field strengths,
the orientation distribution is highly polarized (close to saturation)
and hence requires a dense orientation quadrature to resolve.
(The orientation distribution approaches a $\delta$-function
in the limit of infinite electric field.)

\subsection{Accuracy of water functionals} \label{sec:WaterAccuracy}

Finally, we turn to a comparison of the excess functionals for water
suitable for \emph{ab initio} solvation methods. In particular,
we focus on the scalar-EOS functional of section~\ref{sec:ScalarEOS},
the bonded-voids functional \cite{BondedVoids} and the functional
of Lischner et al. \cite{LischnerH2O}. The last functional is based on experimental correlations
functions, which we will refer to as the `fitted-correlations' functional.
We perform all calculations in one-dimensional planar or radial grids,
using the Fluid1D sub-project of JDFTx \cite{Fluid1D}.
We use the Euler(20) orientation quadrature, with $n_\alpha=1$ to exploit
rotational symmetry in the transverse directions. (See \ref{sec:Discretization1D}.)

\begin{figure}
\includegraphics[width=4.8in]{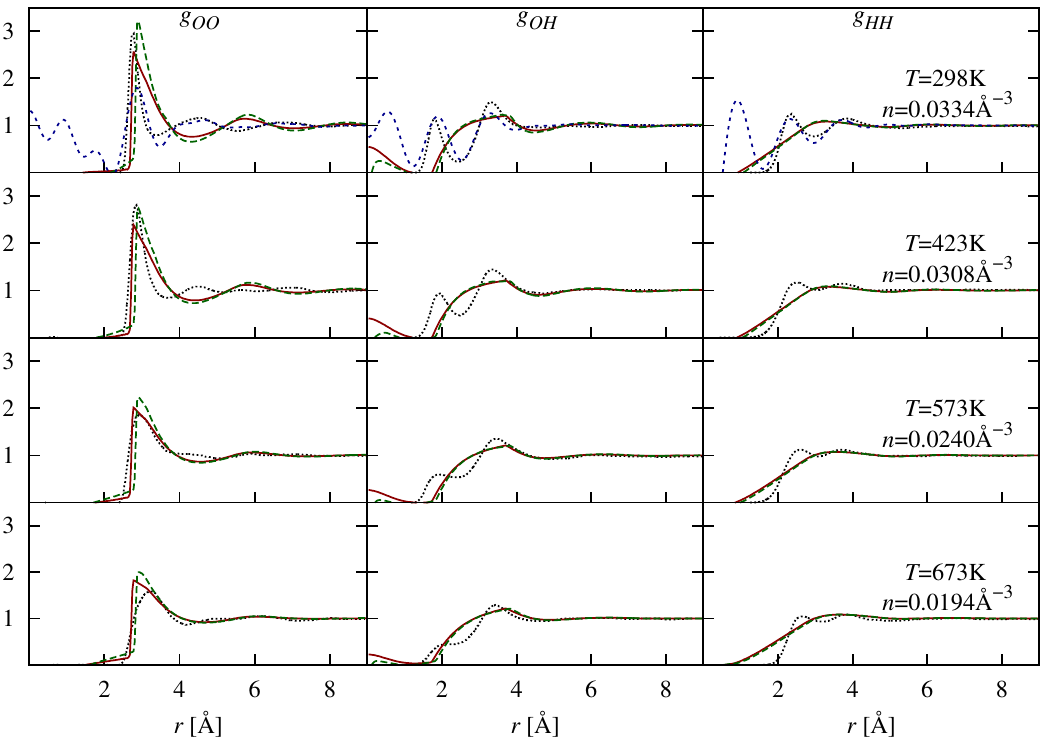}
\caption{Partial radial distributions (site-site correlation functions)
for the scalar-EOS water functional (solid red lines),
bonded-voids functional \cite{BondedVoids} (long-dashed green lines)
and fitted-correlations functional \cite{LischnerH2O} (short-dashed blue lines),
compared to experimental pair correlations of water
from Soper et al. \cite{SoperEPSR} (black dotted lines).
The position and location of the first $g_{OO}$ peak for scalar-EOS and bonded-voids
are in reasonable agreement with experiment, but the remaining structure
resembles that of a close-packed hard sphere fluid rather than a tetrahedrally
bonded one. The fitted-correlations functional is defined only at 298~K
and captures the features of the correlation functions by construction,
but suffers from short-ranged artifacts due to
the bandwidth-limited fitting procedure of \cite{LischnerH2O}.
\label{fig:gXX}}
\end{figure}

First we examine the pair correlation functions $g_{\alpha\beta}$
of the bulk fluid computed using the Ornstein-Zernike relation
for the rigid-molecular fluid which may be written as
\begin{equation}
\tilde{h} = (1-\tilde{I}\tilde{c}\bar{N})^{-1}\tilde{I}\tilde{c}\tilde{I},
\label{eqn:OZ}
\end{equation}
which is a matrix equation in Fourier space for each wave vector $k$.
Here, $\tilde{h}_{\alpha\beta}(k)$ is the Fourier transform of
$g_{\alpha\beta}(r)-1$, $\tilde{I}_{\alpha\beta}(k) = j_0(kR_{\alpha\beta})$
is the intra-molecular structure factor with $R_{\alpha\beta}$ being the
distance between sites $\alpha$ and $\beta$ within the molecule,
$\bar{N}$ is the bulk number density of fluid molecules, and
$\tilde{c}_{\alpha\beta}(k)$ is the Fourier transform of the
direct correlation function $c_{\alpha\beta}(\vec{r}-\vec{r}')
=(-1/T) \delta^2 F\sub{ex}/\delta N_\alpha(\vec{r})\delta N_\beta(\vec{r}')$
evaluated in the limit of the uniform fluid.\footnote{
The relation (\ref{eqn:OZ}) may be generalized to mixtures of rigid-molecular fluids
by replacing $\bar{N}$ with a diagonal matrix with the bulk number density
of each component in the mixture, and setting $\tilde{I}_{\alpha\beta}=0$
when $\alpha$ and $\beta$ belong to different components of the mixture.}

The direct correlation functions are computed analytically in Fourier space
for a set of wave vectors corresponding to the spherical Bessel function
basis with 1024 basis functions and a radial extent $r\sub{max}=64$~bohrs
(see \ref{sec:Discretization1D}), and the pair correlation functions
are computed via (\ref{eqn:OZ}) using numerical spherical Bessel transforms.
Figure~\ref{fig:gXX} compares the pair correlations for all three functionals
under consideration compared against those obtained by Soper et al \cite{SoperEPSR}
from neutron diffraction data by empirical-potential structure refinement (EPSR).

The scalar-EOS functional correctly captures the location and height of the first
peak in $g_{OO}(r)$, but produces a secondary structure reminiscent of the
close-packed coordination of the hard sphere fluid rather than the tetrahedral
coordination exhibited by water. The split hydrogen peaks in the experimental data
are fused into a single broader one with the same particle content.
These are qualitatively the same features as the bonded-voids functional,
but with slightly better agreement for the scalar-EOS functional.
After all, the motivation for the scalar-EOS functional was to simplify
the bonded-voids functional because it captured free energies of cavity formation
reasonably \emph{despite} not exhibiting features of tetrahedral correlation.
The fitted-correlations functional reproduces some of the features of the
experimental correlation functions by construction, but exhibits artifacts
at short distances due to the bandwidth limitation in the fitting procedure
for the correlations (and partly because it does not employ a hard sphere reference).

\begin{figure}
\begin{center}\includegraphics[width=3.4in]{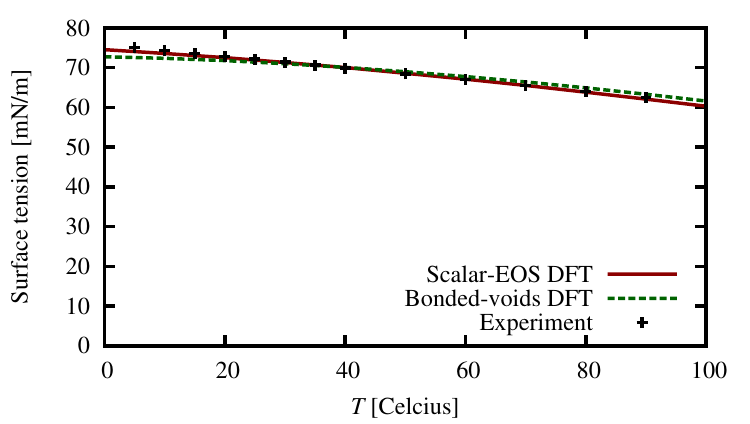}\end{center}
\caption{Energy of the planar vapor-liquid interface for the
scalar-EOS and bonded-voids water functionals as a function of temperature,
compared to the experimental values for surface tension \cite{LangeHandbook}.
Both functionals fit the range parameter of the Lennard-Jones
pair potential to the experimental surface tension at 298~K,
and the scalar-EOS functional reproduces the temperature dependence
more accurately than the bonded-voids one. (The fitted-correlations functional
is omitted from this plot, since it is defined only at  298~K.)
\label{fig:sigmavsT}}
\end{figure}

Next, we examine the free energies of planar liquid-vapor interface for each functional.
The calculations are performed on a one-dimensional planar grid of length 96~bohrs
with 768 sample points and basis functions.
For each temperature, the pressure is adjusted to the boiling point, which corresponds to
equal chemical potentials and bulk grand free energy densities for the two phases.
The hard sphere radius $R\sub{HS} = 1.36$~\AA~ for the scalar-EOS functional
was determined by matching the interface energy obtained from such a calculation
at 298~K to the experimental value for the surface tension 72.0~mN/m.\footnote{
The attraction range parameter $\sigma_U$ in the bonded-voids model \cite{BondedVoids}
and the smoothing parameter $r_0$ of the fitted-correlations model \cite{LischnerH2O}
were also fit to reproduce the surface tension at 298~K using similar calculations.}
Figure~\ref{fig:sigmavsT} compares the temperature dependence of this interface energy
against experimental values for the surface tension. The scalar-EOS functional captures
the trend in the experimental data slightly better than the bonded-voids functional.

\begin{figure}
\includegraphics[width=4.8in]{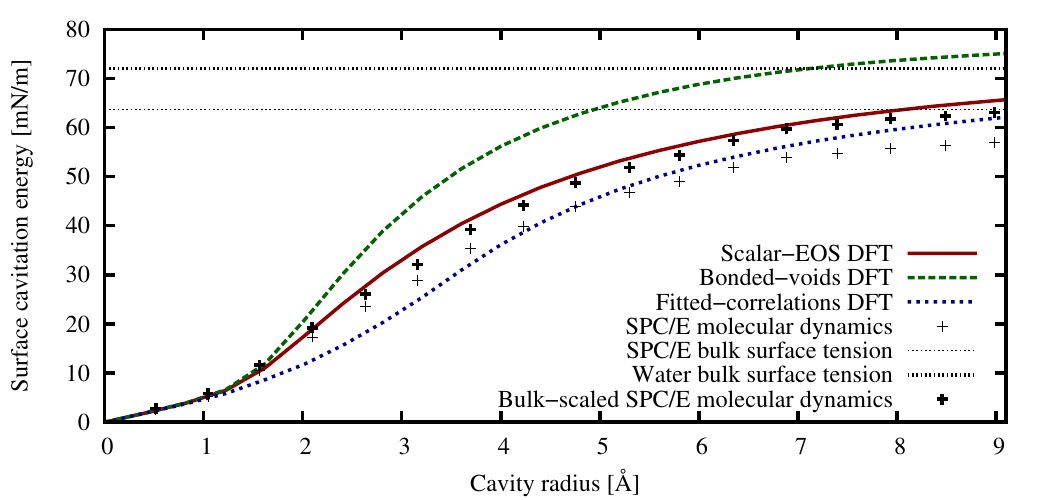}
\caption{Variation of the solvation energy of hard spheres that exclude
the oxygen sites of water from their interior, with the radius of such spheres,
compared to the SPC/E molecular dynamics results of \cite{HardSphereSPCE}.
The SPC/E model underestimates the bulk surface tension of water
by $10\%$ \cite{SurfaceTensionMD}, and we have included a scaled
version of the SPC/E data as a reasonable guess for real water.
The scalar-EOS functional agrees with the bulk-scaled SPC/E data accurately,
while the  fitted-correlations functional systematically underestimates and
the bonded-voids functional overestimates the free-energy of cavity formation.
\label{fig:sigmavsradius}}
\end{figure}

\begin{figure}
\includegraphics[width=4.8in]{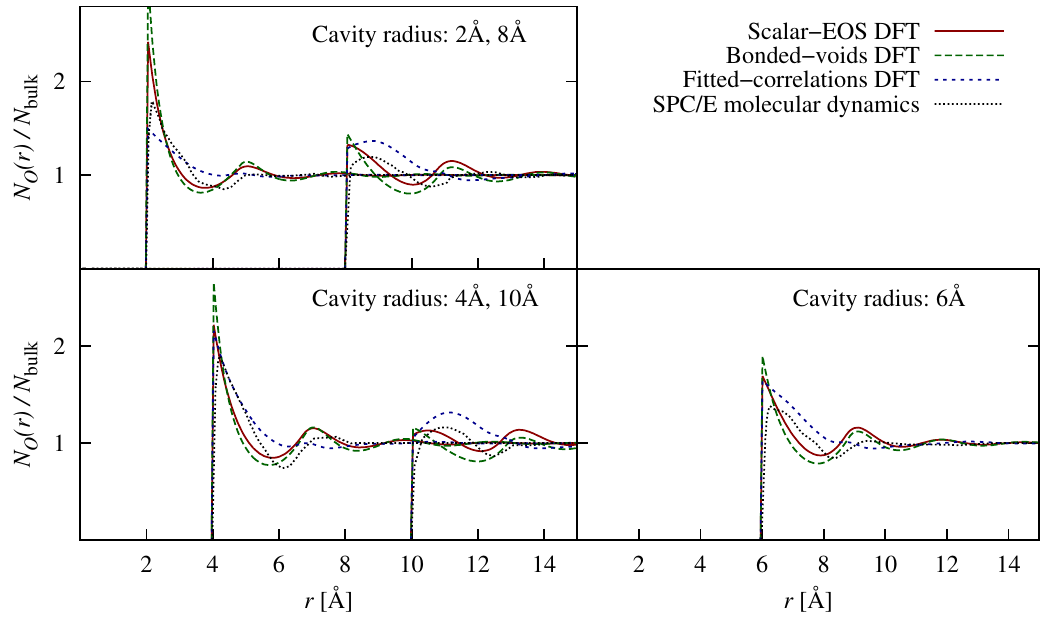}
\caption{Radial density profiles around hard spheres that exclude the oxygen sites
of water from their interior, for spheres of radius 2, 4, 6, 8 and 10~\AA,
compared to the SPC/E molecular dynamics results of \cite{HardSphereSPCE}.
The fitted-correlations functional misses the secondary peaks in the profiles,
while the scalar-EOS and bonded-voids functional overestimate the contact density
and the secondary structure, with best agreement provided by the scalar-EOS model.
\label{fig:hardsphere}}
\end{figure}

The planar interface energies provide a means to calibrate the liquid functionals
against experimental measurements, and the excellent agreement for the temperature
dependence after adjusting the surface tension at one temperature is promising.
However, the applicability of a functional for molecular solvation calculations
depends on its ability to accurately describe the free energies required to form
cavities of molecular dimensions. A standard test case is the solvation free energy
for microscopic hard spheres in the fluid. We compute the cavitation energies for hard spheres
of radii $R$ ranging from 0 to 9~\AA, with external potentials $V_O(r)=(1~E_h)\theta(R-r)$
and $V_H(r)=0$ that exclude the oxygen site of water from the interior of the spheres.
The calculations are performed on a one-dimensional radial grid of extent
$r\sub{max}=64$~bohrs ($\approx34$~\AA) with 512 sample points and basis functions.

Figure~\ref{fig:sigmavsradius} compares the variation of the hard sphere
solvation energy per surface area with sphere radius for all three functionals
with SPC/E molecular dynamics estimates for the same from \cite{HardSphereSPCE}.
For large spheres, the surface curvature effects become negligible and the
surface energy approaches the planar surface tension; whereas for small enough
spheres the cavitation energy is proportional to the volume
($\Delta\Phi = N\sub{bulk}T\times(4\pi R^3/3)$, so that $\Delta\Phi/(4\pi R^2)\propto R$).
Note that all three functionals agree perfectly with the molecular dynamics results
in the small radius limit, and they all approach the bulk experimental
surface tension in the large radius limit (after overshooting the bulk value
in the bonded-voids case). However the SPC/E model underestimates the
bulk surface tension to be 65~mN/m \cite{SurfaceTensionMD}
compared to the experimental value of 72~mN/m, and therefore
the molecular dynamics results for the sphere solvation energies
are also underestimated by a similar amount for the larger spheres.
Consequently, we include the molecular dynamics results scaled up by the
ratio of experimental to SPC/E surface tensions as a reasonable guess
for the hard sphere cavitation energy of \emph{real} water in 
Figure~\ref{fig:sigmavsradius} (in addition to the unscaled values).\footnote{
The TIP4P/2005 pair potential for water captures the bulk surface tension
much more accurately than SPC/E \cite{SurfaceTensionMD}, and it would be
interesting to compare our density functional results to simulations
of microscopic hard sphere solvation with that model. However, such results
for TIP4P/2005 (analogous to \cite{HardSphereSPCE} for SPC/E)
have not yet been published to our knowledge.}
The scalar-EOS functional significantly outperforms bonded-voids
and fitted-correlations in its agreement with the bulk-scaled
molecular dynamics results, and is the best candidate for an accurate
density functional description of cavitation energies in liquid water.

We next examine the distribution of water around these spherical cavities
of selected sizes in Figure~\ref{fig:hardsphere}. As expected from
the results for the free energies, the density profiles of the scalar-EOS
functional are in closest agreement with the SPC/E molecular dynamics
results of \cite{HardSphereSPCE}. The bonded-voids functional overestimates
the structure in the liquid, which is expected since it also overestimated
the structure in the pair correlations (Figure~\ref{fig:gXX}).
The fitted-correlations functional severely underestimates the secondary
structure in the density profiles despite better qualitative agreement
with the experimental pair correlation functions.

\begin{figure}
\begin{center}\includegraphics[width=4.8in]{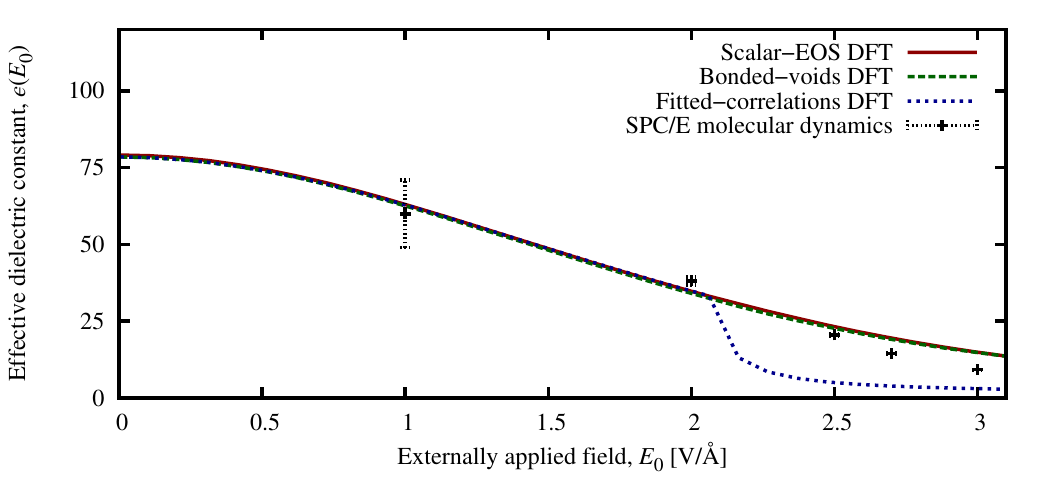}\end{center}
\caption{Nonlinear dielectric response of the water functionals compared
to SPC/E molecular dynamics results from \cite{NonlinearEpsSPCE}.
All three functionals provide essentially the same dielectric response,
as this is determined by a competition between the molecular ideal gas entropy
and the scaled mean field electrostatics. The minor differences arise
from differences in the equations of state due to electrostriction
(change of bulk-density in strong fields). The fitted-correlations functional
has an unphysical instability accompanied by a rapid increase in density
and drop in dielectric constant at an external field $\approx 2$~V/\AA,
due to an underestimation of the compressibility at high pressures
by its polynomial excess free energy density model.
\label{fig:nonlineareps}}
\end{figure}

Finally, we turn to the last key ingredient for a successful theory of solvation:
nonlinear dielectric response. The typical electric fields in the vicinity
of polar molecules are $\sim$~V/\AA~ i.e. $10^{10}$~V/m, which corresponds
to strong non-linearities and significant dielectric saturation.
Here, we examine the nonlinear dielectric constant defined by $\epsilon(E_0) = E_0
/ (E_0 - 4\pi P)$, where $E_0$ is a macroscopic externally applied field
and $P$ is the corresponding bulk polarization density in the liquid.

At equilibrium, a liquid in a macroscopic parallel-plate capacitor adopts
uniform density and polarization except for microscopic regions around the plates.
The free energy of that capacitor is dominated by the bulk; the regions
next to the plates only contribute via long-range interactions
of the bound sheet-charge densities $\pm P$ in the liquid.
Accounting for the interaction of these sheet charges with the external field,
and with each other via the scaled mean-field Coulomb interaction,
we can show that the effective free energy density minimized by the
macroscopic capacitor at equilibrium is
\begin{multline*}
\phi(p_\omega) = T \int \frac{\textrm{d}\omega}{8\pi^2} p_\omega
	\left(\log\frac{p_\omega}{N_\textrm{ref}} - 1\right)
	- \sum_\alpha \mu_\alpha N_\alpha + f\sub{ex}(\{N_\alpha\}) \\
	- \vec{E}_0\cdot\vec{P} + A_\epsilon(T) \frac{4\pi P^2}{2}.
\end{multline*}
Here, $f\sub{ex}$ is the excess-free energy density of the uniform fluid
(which is determined entirely by the equation of state) and
$\vec{P} = \int \frac{\textrm{d}\omega}{8\pi^2} p_\omega \omega\circ\vec{P}\sub{mol}$
is the polarization density, with $\vec{P}\sub{mol}$ being the
dipole moment of the fluid molecule in its reference orientation.
We therefore minimize this free energy density on a planar grid with
a single grid point to obtain the equilibrium $P$ for each applied $E_0$,
thereby avoiding the need for setting up a capacitor in a large simulation cell.

All three functionals considered here employ the same scaled mean-field
electrostatic interaction constrained to produce the bulk dielectric response
as proposed by Lischner et al. \cite{LischnerH2O}. The physics of dielectric
saturation is captured by an interplay of this term with the entropy of the
ideal gas of rigid molecules, which again is common to all three functionals.
Consequently, their nonlinear dielectric response is very similar and compares
quite well with the SPC/E molecular dynamics results \cite{NonlinearEpsSPCE}
as shown in Figure~\ref{fig:nonlineareps}. The minor differences between
the functionals are due to the different uniform fluid excess free energy
densities ($f\sub{ex}$) which correspond to different approximations to the
equation of state of the fluid. The fitted-correlations functional employs
a polynomial model for $f\sub{ex}$ obtained from the bulk modulus and its pressure
derivative at ambient conditions \cite{LischnerH2O}, which underestimates
the bulk modulus at high compression. This causes the instability at high fields
associated with a rapid increase in density, seen as a drop in the
dielectric response at $\approx2$~V/\AA~ in Figure~\ref{fig:nonlineareps}.

\section{Conclusions}

We construct a general framework for the classical density-functional theory
of rigid-molecular fluids that avoids the inversion problem associated with
site-density constraints by switching to the orientation density as the key variable.
We show that the independent variables in previous solutions, such as
ideal-gas effective site potentials, are compressed maximum-entropy representations
of the orientation density. We then motivate other representations with superior
convergence properties which are variational approximations to the free energy.
The self-representation, directly minimizing over the orientation density $\{p_\omega\}$,
exhibits the fastest convergence for conjugate gradients minimization, but requires
memory in proportion to the size of the quadrature for orientation integrals.
The site-potential representation $\{\psi_\alpha\}$, although exact in principle,
is impractical due to poor convergence, particularly in the presence of strong electric fields.
We introduce the multipole representation $\{\mu\vec{\epsilon}\}$ which exhibits
comparable convergence to the self-representation without the memory overhead,
is effectively more accurate than $\{\psi_\alpha\}$ at practical iteration counts
despite being a (variational) approximation, and enables efficient large-scale
\emph{ab initio} solvation in polar molecular fluids within the framework of
joint density-functional theory.

We extend the algebraic formulation of electronic density-functional theory,
DFT++ \cite{AlgebraicDFT}, and present the discretization of our general framework
and excess functionals for practical calculations in a basis-independent manner.
The methods developed in this paper form the basis for the fluid sector of the
open-source electronic density-functional theory software JDFTx \cite{JDFTx},
which provides a three-dimensional plane-wave basis implementation of this work.
Additionally, a one dimensional version implementing the three basis sets of
\ref{sec:Discretization1D} is distributed as a sub-project of JDFTx \cite{Fluid1D},
suitable for rapid prototyping and development of fluid functionals within this framework.

In addition to the general framework for polar fluids,
we construct a practical free energy functional for liquid water
which improves on the accuracy of earlier functionals,
the bonded-voids model \cite{BondedVoids} based on Wertheim perturbation
and the fitted-correlations model \cite{LischnerH2O}
based on experimental correlation functions of water,
while minimizing complexity and avoiding over-parametrization.
We show that this `scalar-EOS' functional accurately captures the key quantities
of interest for \emph{ab initio} solvation calculations: free energies for formation of
microscopic cavities in the fluid, and non-linear dielectric response.
Within joint density-functional theory, the methods developed here
provide an accurate and efficient description of solvent environments,
thereby enabling a focused electronic structure study of solvated
biological and chemical systems of technological relevance.

This work was supported as a part of the Energy Materials Center at Cornell (EMC$^2$),
an Energy Frontier Research Center funded by the U.S. Department of Energy,
Office of Science, Office of Basic Energy Sciences under Award Number DE-SC0001086.

\appendix
\section{Efficient quadratures for orientation integrals} \label{sec:SO3quad}

Efficient discretization of the orientation integrals is critical to the performance
of any of the representations of Section~\ref{sec:Representations}
and determines the very practicality of the $p_\omega$ (self) representation.
Here, we list efficient quadratures for discretizing integrals over $\omega$,
$\int \frac{\textrm{d}\omega}{8\pi^2} f(\omega) \to \sum_i W_i f(\omega_i)$.

The simplest approach is to label orientations by ZYZ-Euler angles
$\omega=(\alpha,\beta,\gamma)$ and use the outer product of
a Gauss-Legendre quadrature for $\beta\in[0,\pi]$
and Gauss-Fourier quadratures for the periodic $\alpha,\gamma\in[0,2\pi)$.
More efficient quadratures may be constructed as an outer product
using the $\mathbb{S}_2\times\mathbb{S}_1$ structure of $SO(3)$,
or by working directly on $SO(3)$ without an outer product structure \cite{SphereSampling}.

In \cite{SphereSampling}, quadratures on $SO(3)$ are optimized to minimize the RMS
error in the integrals of all $D^j_{m_1m_2}(\omega)$ up to some $j\sub{max}$.
We focus on quadratures that are \emph{exact} up to some
$j\sub{max}$,
\begin{equation}
\sum_i W_i D^j_{m_1m_2}(\omega_i)
	= \sum_i W_i d^j_{m_1m_2}(\beta_i) \textrm{e}^{i(m_1\alpha_i+m_2\gamma_i)}
	~=~ \delta_{j0} \label{eqn:SO3quadExactness}
\end{equation}
for all $|m_1|,|m_2| \le j \le j\sub{max}$,
and can be optimized further using the symmetry of the molecule at hand.
For simplicity, we only consider $\mathbb{Z}_n$ symmetry about a single axis,
chosen to be the $z$-axis of the molecule frame without loss of generality.
The quadratures considered then fall into 3 classes:
\begin{enumerate}[(a)]
	\item Symmetry groups of Platonic solids \cite{SphereSampling}
	\item Outer products of a spherical $j$-design \cite{SphericalDesignLib}
		on $\mathbb{S}_2(\alpha,\beta)$ with a uniform quadrature on $\mathbb{S}_1(\gamma)$
	\item Outer product quadrature on all 3 Euler angles $\alpha$, $\beta$ and $\gamma$.
\end{enumerate}
Each of these these classes consists of uniformly spaced nodes of equal weights in $\gamma$
for each $(\alpha,\beta)$. Grouping the nodes as $(\alpha_k,\beta_k,\gamma_k+2n\pi/n_\gamma)$
for $n\in0,\ldots,n_\gamma-1$ with total weight $W_k$ for each group,
(\ref{eqn:SO3quadExactness}) can be reduced to
\begin{equation}
\sum_k W_k d^j_{m_1m_2}(\beta_k) \textrm{e}^{i(m_1\alpha_k+m_2\gamma_k)}
	= \delta_{j0} \label{eqn:SO3quadExactnessUniform}
\end{equation}
for all $|m_1|,|m_2| \le j \le j\sub{max}$ such that $m_2$ is a multiple
of $n_\gamma$. Therefore if $n_\gamma > j\sub{max}$
(which is the case for all but the Icosahedron rotation group),
(\ref{eqn:SO3quadExactnessUniform}) further simplifies to
\begin{equation}
\sqrt{4\pi} \sum_k W_k Y^j_{m}(\beta_k,\alpha_k)
	= \delta_{j0} \label{eqn:S2quadExactness}
\end{equation}
for all $|m| \le j \le j\sub{max}$ using the relations of $D^j_{m0}$
to the spherical harmonics.

A spherical $j\sub{max}$-design is a set of points on the unit sphere
that satisfies (\ref{eqn:S2quadExactness}) with uniform weights $W_k$,
and hence it yields an $SO(3)$ quadrature exact to $j\sub{max}$
when combined with a uniform quadrature with $j\sub{max}+1$
nodes on $\mathbb{S}_1(\gamma)$. We use the spherical designs
with the smallest number of nodes for each $7 \le j\sub{max} \le 21$
tabulated in \cite{SphericalDesignLib} to form the quadratures of class (b).
The quadratures of lower order reduce to class (a), specifically
the rotation groups of the Tetrahedron at $j\sub{max}=2$,
Octahedron at $j\sub{max}=3$ and Icosahedron at $j\sub{max}=5$.

The Gauss-Legendre quadrature with $n_\beta$ nodes on $\cos\beta\in[-1,1]$
is exact for the integration of all polynomials up to order $2 n_\beta-1$.
The outer product of this with a uniform quadrature with $2 n_\beta$
nodes on $\alpha\in[0,2\pi)$ satisfies (\ref{eqn:S2quadExactness})
for $j\sub{max}=2 n_\beta-1$, and hence also (\ref{eqn:SO3quadExactness})
to that order when combined with $2 n_\beta$ uniform samples on $\gamma$.

Finally the reduction by $\mathbb{Z}_n$ symmetry about the $z$-axis in
the molecule frame amounts to replacing $\mathbb{S}_1(\gamma)$
with $\mathbb{S}_1/\mathbb{Z}_n$. This is achieved by a uniform sampling
of $\lceil n_\gamma/n \rceil$ points on $\gamma\in[0,2\pi/n)$,
which retains the exactness to $j\sub{max}$ for functions
with this symmetry with a reduction of up to $n$ in the number of nodes required.

The accuracy of these quadratures for the classical density functional
theory of rigid molecules is explored in section \ref{sec:Convergence}.
The quadratures considered there are listed in Table \ref{tab:SO3quad}
along with their $j\sub{max}$, the number of nodes for sampling
$SO(3)/\mathbb{Z}_n$ in general and $SO(3)/\mathbb{Z}_2$ in particular,
which is the case relevant for water.
Note that the Euler quadrature with $n_\beta=3$ needs almost twice
as many nodes as the Icosahedron group for the same $j\sub{max}=5$,
but the relative inefficiency of the Euler quadratures decreases
with $j\sub{max}$ and becomes less than $1\%$ between the
$n_\beta=11$ Euler quadrature and the 21-design at $j\sub{max}=21$.

\begin{table}
\begin{center}
\begin{tabular}{| c | c | c | c |}
\hline
\multirow{2}{*}{} & \multirow{2}{*}{$j\sub{max}$} & \multicolumn{2}{c|}{Number of quadrature nodes for} \\
\cline{3-4} & & $SO(3)/\mathbb{Z}_n$ & $SO(3)/\mathbb{Z}_2$ \\
\hline
Tetrahedron & 2 & $4\times\lceil3/n\rceil$ & 8 \\
Octahedron & 3 & $6\times\lceil4/n\rceil$ & 12 \\
Icosahedron & 5 & $12\times\lceil5/n\rceil$ & 36 \\
\hline
 7-design &  7 & $ 24\times\lceil 8/n\rceil$ &   96 \\
 8-design &  8 & $ 36\times\lceil 9/n\rceil$ &  180 \\
 9-design &  9 & $ 48\times\lceil10/n\rceil$ &  240 \\
10-design & 10 & $ 60\times\lceil11/n\rceil$ &  360 \\
11-design & 11 & $ 70\times\lceil12/n\rceil$ &  420 \\
12-design & 12 & $ 84\times\lceil13/n\rceil$ &  588 \\
13-design & 13 & $ 94\times\lceil14/n\rceil$ &  658 \\
14-design & 14 & $108\times\lceil15/n\rceil$ &  864 \\
15-design & 15 & $120\times\lceil16/n\rceil$ &  960 \\
16-design & 16 & $144\times\lceil17/n\rceil$ & 1296 \\
17-design & 17 & $156\times\lceil18/n\rceil$ & 1404 \\
18-design & 18 & $180\times\lceil19/n\rceil$ & 1800 \\
19-design & 19 & $204\times\lceil20/n\rceil$ & 2040 \\
20-design & 20 & $216\times\lceil21/n\rceil$ & 2376 \\
21-design & 21 & $240\times\lceil22/n\rceil$ & 2640 \\
\hline
Euler$(n_\beta)$ & $2 n_\beta-1$ & $2n_\beta^2 \times\lceil2n_\beta/n\rceil$ & $2n_\beta^3$ \\
\hline
\end{tabular}
\end{center}
\caption{List of explored quadratures, their degree of exactness $j\sub{max}$,
and the number of nodes in sampling $SO(3)/\mathbb{Z}_n$. The Euler angles
corresponding to the platonic solid rotation groups are listed in \cite{SphereSampling}.
The $j$-designs are constructed as an outer product of the spherical $j$-designs
with fewest points for each $j$ from \cite{SphericalDesignLib} used for $(\alpha,\beta)$
with $\lceil (j+1)/n \rceil$ uniform samples on $\gamma\in[0,2\pi/n)$.
Each Euler$(n_\beta)$ quadrature is an outer product of a
$n_\beta$-point Gauss-Legendre quadrature on $\cos\beta\in[-1,1]$, 
a uniform $2 n_\beta$-point quadrature on $\alpha\in[0,2\pi)$,
and a uniform $\lceil 2 n_\beta/n \rceil$-point quadrature on $\gamma\in[0,2\pi/n)$.
\label{tab:SO3quad}}
\end{table}

\section{One-dimensional discretization for special geometries} \label{sec:Discretization1D}

The discretization of three-dimensional space according to
Section~\ref{sec:Discretization}, along with the orientation quadratures
of \ref{sec:SO3quad} provide a practical route to computations with the
rigid-molecular classical density functional framework of
Section~\ref{sec:IdealGas} in arbitrary geometries and basis sets. 
However, the development and testing of new excess functionals for liquids
primarily require calculations in high-symmetry configurations.
Here, we detail the formulation of highly-efficient discretizations
of planar, cylindrical and spherical geometries on a one-dimensional grid,
which allow for the rapid prototyping of excess functionals
employed in Section~\ref{sec:WaterAccuracy} and \cite{BondedVoids}.

The discretization of space is performed in the framework of
Section~\ref{sec:Discretization}, but with special basis sets
exploiting the symmetry. The three geometries we consider here
are
\begin{enumerate}
\item Planar, where all spatial dependence is along $z$,
\item Cylindrical, with dependence only on the distance from the $z$-axis $\rho$, and
\item Spherical, with dependence only on distance from origin $r$.
\end{enumerate}
Each of these geometries require only a one-dimensional discretization.
For the planar geometry, we impose mirror-symmetry boundary conditions
at the ends of the grid, and pick a basis of cosines and a corresponding
quadrature grid suited for the Discrete Cosine Transform \cite{DCT}.
For the spherical and cylindrical geometries, we impose Neumann
boundary conditions at some maximum radius, and choose a finite basis
of spherical and cylindrical Bessel functions respectively, along with
a quadrature grid suited for the Discrete Bessel Transform \cite{BesselTransform}
.\footnote{The Discrete Bessel Transform of \cite{BesselTransform} is based on
Dirichlet boundary conditions; the extension of that approach to Neumann boundary
conditions is straightforward, and the results are summarized in Table~\ref{tab:Basis1D}.}
The definition of the basis functions, quadrature grid and the matrix elements
for the operators of Section~\ref{sec:Discretization} are summarized in
Table~\ref{tab:Basis1D}.

\def\bksp{\hspace*{-1.1ex}} 

\begin{table}
\begin{center}
\begin{tabular}{| m{0.7in} | m{1.15in} | m{1.05in} | m{1.1in} |}
\hline
	& Planar
	& Cylindrical
	& Spherical
\\\hline
Coordinate\allowbreak System
	& $(x,y,z)$
	& $(\rho,\phi,z)$
	& $(r,\theta,\phi)$
\\\hline
Symmetry
	& $f(\vec{r}) \to f(z)$
	& $f(\vec{r}) \to f(\rho)$
	& $f(\vec{r}) \to f(r)$
\\\hline
Boundary\allowbreak conditions
	& $f'(0) = f'(L) = 0$
	& $f'(\rho\sub{max}) = 0$
	& $f'(r\sub{max}) = 0$
\\\hline
Basis $b_i(\vec{r})$
	&\bksp$\begin{array}{l}
		\tilde{w}_i \cos(G_i z), \\
		G_i = i\pi/L,\\
		\tilde{w}_i = \frac{2}{(1+\delta_{i0})L}
	\end{array}$
	&\bksp$\begin{array}{l}
		\tilde{w}_i J_0(G_i\rho), \\
		G_i = Y_i/\rho\sub{max}, \\
		\tilde{w}_i = \frac{J_0^{-2}(Y_i)}{\pi\rho\sub{max}^2}
	\end{array}$
	&\bksp$\begin{array}{l}
		\tilde{w}_i j_0(G_i r), \\
		G_i=y_i/r\sub{max}, \\
		\tilde{w}_i = \frac{j_0^{-2}(y_i)}{\left(2-\frac{2}{3}\delta_{i0}\right)\pi r\sub{max}^3}
	\end{array}$
\\\hline
Quadrature\allowbreak grid $\{\vec{r}_j\}$
	& $z_j = (j+\frac{1}{2})\frac{L}{S}$
	& $\rho_j = X_{j+1}\frac{\rho\sub{max}}{Y_S}$
	& $r_j = x_{j+1}\frac{r\sub{max}}{y_S}$
\\\hline
$\mathcal{I}_{ji}$
	& $\tilde{w}_i \cos\left( (j+\frac{1}{2})\pi\frac{i}{S} \right)$
	& $\tilde{w}_i J_0\left( X_{j+1}\frac{Y_i}{Y_S} \right)$
	& $\tilde{w}_i j_0\left( x_{j+1}\frac{y_i}{y_S} \right)$
\\\hline
$\mathcal{J}_{ij}$
	&\bksp$\begin{array}{l}
		w_j \cos\left( (j+\frac{1}{2})\pi\frac{i}{S} \right), \\
		w_j = \frac{L}{S}
	\end{array}$
	&\bksp$\begin{array}{l}
		w_j J_0\left( X_{j+1}\frac{Y_i}{Y_S} \right), \\
		w_j = \frac{4\pi\rho\sub{max}^2}{Y_S^2 J_1^2(X_{j+1})}
	\end{array}$
	&\bksp$\begin{array}{l}
		w_j j_0\left( x_{j+1}\frac{y_i}{y_S} \right), \\
		w_j = \frac{4\pi^2r\sub{max}^3}{y_S^3 j_1^2(x_{j+1})}
	\end{array}$
\\\hline
$\mathcal{O}_{i'i}$ & \multicolumn{3}{c|}{$\tilde{w}_i \delta_{i'i}$} \\\hline
$\mathcal{L}_{i'i}$ & \multicolumn{3}{c|}{$-G_i^2 \tilde{w}_i \delta_{i'i}$} \\\hline
$(\mathcal{J^\dag OJ})_{j'j}$ & \multicolumn{3}{c|}{$w_j \delta_{j'j}$} \\\hline
$(g(r)\ast)_{i'i}$ & \multicolumn{3}{c|}{$\delta_{i'i}\int4\pi r^2\textrm{d}r g(r) j_0(G_ir)$} \\\hline
\end{tabular}
\end{center}
\caption{Definition of the basis functions for the high-symmetry geometries
- planar, cylindrical and spherical - with one-dimensional discretizations
of sample count $S$, and matrix elements of the operators of
Section~\ref{sec:Discretization} for each of these basis sets.
The basis functions are labeled by $i=0,1,\cdots,S-1$ for each basis set,
and $X_i$, $x_i$, $Y_i$ and $y_i$, are the $i\super{th}$ roots
of $J_0(x)$, $j_0(x)$, $J_0'(x)$ and $j_0'(x)$ respectively, with $Y_0 = y_0 \equiv 0$.
The quadrature grid has the same number of points $S$ as the basis size,
and are labeled by $j=0,1,\cdots,S-1$.
\label{tab:Basis1D}}
\end{table}

All three basis sets are derived from the eigenfunctions of the three-dimensional
Laplace equation in various geometries, and are therefore intricately linked
to the three-dimensional plane-wave basis: the basis functions are indexed
by $G_i$, the magnitude of the corresponding plane-wave momentum.
Consequently, the Laplacian and convolutions by spherical functions
are diagonal in these basis sets as well, as indicated in Table~\ref{tab:Basis1D}.
The transform operators $\mathcal{I}$ and $\mathcal{J}$ reduce to the
`DCT type III' and `DCT Type II' fast Fourier transforms \cite{FFTW3} respectively
in the planar geometry (or `IDCT' and `DCT' in the notation of \cite{DCT});
the cylindrical and spherical transforms lack an analogous $\mathcal{O}(S\log S)$ method
and are implemented as matrix-vector multiplies with a precomputed Bessel function matrix.

The basis-independent discretization of the scalar-EOS excess functional
(\ref{eqn:DiscreteFex-ScalarEOS}), and site-density excess functionals in general,
carries over to the planar, cylindrical and spherical geometries without modification.
The discretization of the rigid-molecular ideal gas free energy and the
generation of site-densities from independent variables carries over unmodified
for the planar geometry, but is slightly complicated for the cylindrical and
spherical geometries by the fact that the translation operator breaks the symmetry
of the basis set and does not have a one-dimensional representation.

We can however compute the site-densities using (\ref{eqn:DiscreteNalphaFromPomega})
and the orientation-density in the site-potential representation using
(\ref{eqn:DiscretePomegaFromPsiAlpha}) for these basis sets as well,
with minor modifications to the translation operators in those equations.
First, we pick a covariant reference orientation for the molecule,
(relative to the local coordinate frame $(\hat\rho,\hat\phi,\hat z)$
or $(\hat r,\hat\theta,\hat \phi)$), so that $p_\omega(\vec{r})$ is invariant
under the cylindrical or spherical symmetry for each $\omega$ and permits a
one-dimensional representation.\footnote{If we used an invariant reference orientation
as in the three-dimensional case, $p_\omega(\vec{r})$ would be covariant under the symmetry,
so that the spatial dependence of $p_\omega(\vec{r})$ for each $\omega$
would not be cylindrically or spherically symmetric,
and would therefore lack a one-dimensional representation.}
Consequently, the translations involved in (\ref{eqn:DiscreteNalphaFromPomega})
and (\ref{eqn:DiscretePomegaFromPsiAlpha}) would be relative to the local
coordinate frame as well, and hence position-dependent; we therefore need
to generalize the translation operators $\mathcal{T}_{\vec{a}}$
to `warp' operators $\mathcal{T}_{\vec{a}(\vec{r})}$ defined by
$\mathcal{T}_{\vec{a}(\vec{r})} f(\vec{r}) = f(\vec{r} + \vec{a}(\vec{r}))$.
It can be shown that the expressions of Section~\ref{sec:Discretization}
remain valid without modification upon this generalization.

The translation operator for the planar basis is a simple one-dimensional
restriction of its three-dimensional counterpart, and it generalizes to
\begin{equation}
\mathcal{T}_{\vec{a}(\rho)} f(\rho) =
	f\left(\sqrt{(\rho+\vec{a}\cdot\hat{\rho})^2 + (\vec{a}\cdot\hat{\phi})^2}\right)
	\label{eqn:WarpOperatorCylindrical}
\end{equation}
for the cylindrical basis with $f(\rho)\equiv f(2\rho\sub{max}-\rho)$
for $\rho>\rho\sub{max}$, and
\begin{equation}
\mathcal{T}_{\vec{a}(r)} f(r) = f\left(\sqrt{r^2 + a^2 + 2r\vec{a}\cdot\hat{r}}\right)
	\label{eqn:WarpOperatorSpherical}
\end{equation}
for the spherical basis with $f(r)\equiv f(2r\sub{max}-r)$ for $r>r\sub{max}$.\footnote{
The covariant reference frame ensures that $\vec{a}\cdot\hat{\rho}$ and
$\vec{a}\cdot\hat{\phi}$ depend only on $\rho$ (and not $\phi$ and $z$),
and that $\vec{a}\cdot\hat{r}$ depends only on $r$.}
We could compute the matrix elements of these operators in the Bessel basis
and apply the translation as a dense-matrix multiply in basis space,
but those suffer from Nyquist frequency ringing problems similar to
their three-dimensional counterparts. Instead, we compute these operators
in real space using approximate sampling operators
$\mathcal{S}_{\vec{a}(\vec{r})}$ based on constant or linear-spline
interpolation which preserve non-negativity of scalar fields.
 
The results for the scalar-EOS water functional in Section~\ref{sec:WaterAccuracy}
and the bonded-voids water functional in \cite{BondedVoids} were computed
using the discretization scheme of Section~\ref{sec:Discretization},
in the planar and spherical bases, with the warp operator $\mathcal{S}$
computed using linear-spline interpolation as discussed above.
The planar and spherical bases have an additional rotational symmetry
about the local $\hat{z}$ and $\hat{r}$ axes respectively at any point in space
which renders the integral over Euler angle $\alpha$ trivial, so that a quadrature
on $\mathbb{S}_2(\gamma,\beta)$ with no $\alpha$ sampling suffices;
the one-dimensional calculations employ this additional optimization
by using the Euler($n_\beta$) quadratures of \ref{sec:SO3quad},
but with $n_\alpha=1$ irrespective of $n_\beta$.

\bibliographystyle{elsarticle-num-names}

\end{document}